\documentclass[aps, prd, superscriptaddress, preprintnumbers, floatfix, longbibliography, twocolumn, 10pt]{revtex4-2}

\usepackage{graphicx,amsfonts,amsmath,amssymb,amstext}
\usepackage{float,wrapfig}
\usepackage{subfigure, psfrag}
\usepackage{dsfont}
\usepackage{color}
\usepackage[colorlinks=true,urlcolor=blue,citecolor=blue,linkcolor=blue]{hyperref}
\usepackage{bm}

\newcommand{\be}{\begin{equation}}
\newcommand{\ee}{\end{equation}}
\newcommand{\ba}{\begin{eqnarray}}
\newcommand{\ea}{\end{eqnarray}}

\definecolor{purple}{rgb}{0.8,0,0.6}
\definecolor{darkgreen}{rgb}{0.00,0.6,0.00}

\extrafloats{100}

\begin{document}

\title{Anomalous sound attenuation in Weyl semimetals in magnetic and pseudomagnetic fields}
\date{September 2, 2023}

\author{P.~O.~Sukhachov}
\email{pavlo.sukhachov@yale.edu}
\affiliation{Department of Physics, Yale University, New Haven, CT 06520, USA}

\author{L.~I.~Glazman}
\affiliation{Department of Physics, Yale University, New Haven, CT 06520, USA}

\begin{abstract}
We evaluate the sound attenuation in a Weyl semimetal subject to a magnetic field or a pseudomagnetic field associated with a strain. Due to the interplay of intra- and inter-node scattering processes as well as screening, the fields generically reduce the sound absorption. A nontrivial dependence on the relative direction of the magnetic field and the sound wave vector, i.e., the magnetic sound dichroism, can occur in materials with nonsymmetric Weyl nodes (e.g., different Fermi velocities and/or relaxation times). It is found that the sound dichroism in Weyl materials can also be activated by an external strain-induced pseudomagnetic field. In view of the dependence on the field direction, the dichroism may lead to a weak enhancement of the sound attenuation compared with its value at vanishing fields.
\end{abstract}

\maketitle

\section{Introduction}
\label{sec:Introduction}

Weyl and Dirac semimetals are novel materials with unusual electronic properties related to their relativistic-like energy spectrum and nontrivial topology~\cite{Franz:book-2013,Wehling-Balatsky:rev-2014,Felser:rev-2017,Hasan-Huang:rev-2017,Burkov:rev-2018,Armitage:rev-2018,GMSS:book}. Conduction and valence bands in Dirac semimetals touch at Dirac points. Each Dirac point can be viewed as a composition of two Weyl nodes of opposite topological charges or chiralities. The Weyl nodes in Weyl semimetals are separated in momentum and/or energy spaces. Since the total topological charge in the system vanishes, the Weyl nodes always appear in pairs of opposite chiralities~\cite{Nielsen:1981a,Nielsen:1981b,Nielsen:1981c}. One of the most remarkable properties of Dirac and Weyl semimetals is their ability to reproduce various effects previously accessible only in high energy physics~\footnote{A notable exception is the superfluid ${}^3$He-A, where the presence of the chiral anomaly and the applicability of the Adler-Bell-Jackiw equation~\cite{Adler:1969,BJ:1969} were demonstrated~\cite{Bevan-Volovik:1997}} including the celebrated chiral anomaly~\cite{Adler:1969,BJ:1969}.

In essence, the chiral anomaly is the violation of the classical conservation law of chiral or imbalance charge in quantum theory when fermions interact with electromagnetic fields. This process is often referred to as the chirality pumping between the Weyl nodes of opposite chiralities. In the high-energy physics context, the anomaly is crucial for the description of the neutral pion decay into two photons. The chiral anomaly plays an important role in condensed matter physics too. For example, it leads to the ``negative" longitudinal magnetoresistance observed in Dirac (Na$_3$Bi, Cd$_3$As$_2$, and ZrTe$_5$) and Weyl (transition metal monopnictides TaAs, NbAs, TaP, and NbP) semimetals (see Refs.~\cite{Hosur-Qi:rev-2013,Burkov:rev-2015,Gorbar:2017lnp,Hu-Mao:rev-2019,Ong-Liang:rev-2020} for reviews on anomalous transport properties). In these materials, the resistivity decreases with a magnetic field if an electric current is driven along the field.

In addition to the anomalous transport in conventional electromagnetic fields, Dirac and Weyl semimetals allow one to realize unusual chirality-dependent pseudo-electromagnetic or axial gauge fields~\cite{Zhou-Shi:2013,Cortijo-Vozmediano:2015,Grushin:2016,Pikulin-Franz:2016} (see also Ref.~\cite{Ilan-Pikulin:rev-2019} for a review). Unlike electromagnetic fields, the pseudo-electromagnetic ones act on the fermions of opposite chiralities with different sign. Pseudo-electromagnetic fields allow for several interesting effects such as anomalous transport~\cite{Landsteiner:2014,Chernodub-Vozmediano:2014,Cortijo-Vozmediano:2016,Grushin:2016,Pikulin-Franz:2016,Chernodub-Zubkov:2017,Huang:2017,Ferreiros-Bardarson:2018,Behrends-Bardarson:2019}, quantum oscillations~\cite{Pikulin-Franz:2017}, and collective excitations~\cite{Gorbar:2016ygi,Gorbar:2016sey,Gorbar:2016vvg,Gorbar:2017cwv,Chernodub-Vozmediano:2019}, to name a few. Of particular interest are effects related to dynamical deformations, which, for example, could be induced by sound. Such effects include the sound attenuation~\cite{Spivak:2016,Pikulin-Franz:2016,Sengupta-Garate:2020,Pesin-Ilan:2020}, the acoustogalvanic effect~\cite{Sukhachov-Rostami:2019}, the torsion-induced chiral magnetic effect current~\cite{Gao-Kharzeev:2020}, and the axial magnetoelectric effect~\cite{Liang-Balatsky:2020}. It is worth noting that not only deformations affect electron quasiparticles, but also phonons could receive feedback from electrons under certain conditions. In particular, the chiral anomaly for fermions is manifested in phonon dynamics~\cite{Song-Dai:2016,Rinkel-Garate:2017,Rinkel-Garate:2019,Yuan-Xiu:2019}.

While the sound attenuation in Weyl semimetals was already investigated before, the corresponding analysis is incomplete. For example, Ref.~\cite{Spivak:2016} captures only the effect of the anomaly in a narrow parametric regime. In particular, the background contribution to the attenuation coefficient in the absence of magnetic fields, which is similar to the case of conventional multivalley semiconductors~\cite{Weinreich-White:1959,Gurevich-Efros:1963,Gantsevich-Gurevich:1967}, was not taken into account. Furthermore, the effects of the intra-node scattering and the electrostatic screening associated with the inter-node electron dynamics were also neglected. Indeed, while it is common to screen the regular deformation potential~\cite{Akhiezer-Liubarskii:1957,Galperin-Kozub:rev-1979,Abrikosov:book-1988}, it was assumed that the axial or chiral one, which has opposite signs at the Weyl nodes of opposite chiralities, is insensitive to the screening since it does not lead to electric charge deviations. We show that, while this is true in the absence of magnetic fields, the chiral anomaly eventually allows for the oscillating electric charge even for the chiral deformation potential. Therefore, the electric screening plays the crucial role and should be taken into account. This provided one of the main motivations for the present work.

By using the chiral kinetic theory (CKT)~\cite{Xiao-Niu:rev-2010,Son:2013,Stephanov:2012,Son-Spivak:2013}, we calculate the attenuation coefficient in an effective low-energy model of Weyl semimetals subject to a scalar chirality-dependent deformation potential as well as external magnetic and strain-induced pseudomagnetic fields. According to Refs.~\cite{Rinkel-Garate:2017,Song-Dai:2016,Rinkel-Garate:2019}, such a deformation potential could originate from the coupling to a pseudoscalar phonon. Indeed, the conventional momentum-independent deformation potential is strongly screened allowing only for the valley-sensitive part to survive. In the absence of magnetic fields, our final expression for the sound attenuation coefficient agrees with that for usual multivalley semiconductors~\cite{Weinreich-White:1959,Gurevich-Efros:1963,Gantsevich-Gurevich:1967}. Therefore, to activate nontrivial features of Weyl semimetals, one needs to apply external fields. Due to the chiral anomaly, the electrostatic screening, and the intra-node scattering, the sound attenuation becomes \emph{suppressed} in a magnetic field. This suppression is monotonic for symmetric Weyl nodes, which are characterized by the same parameters, and could reach several percent for sufficiently strong fields. On the other hand, in the case where the Weyl nodes have different Fermi velocities or are characterized by different relaxation times, the dependence on the magnetic field becomes nonmonotonic and \emph{magnetic sound dichroism} can be realized. Due to this effect, the attenuation is different for the sound propagating along or opposite to the magnetic field. A similar nonmonotonic and directional dependence occurs also if an external pseudomagnetic field is applied to the semimetal. We notice that \emph{pseudomagnetic sound dichroism} appears even if the Weyl nodes are symmetric. For realistic model parameters, sound dichroism is weak. Its value for optimal (pseudo)magnetic fields is about a few percent of the attenuation coefficient at zero fields.

Recently, a study addressing the sound attenuation in Weyl semimetals in a magnetic field~\cite{Pesin-Ilan:2020} appeared. The authors of Ref.~\cite{Pesin-Ilan:2020} took a somewhat different route in deriving their results, but there is a partial overlap with our findings. Specifically, the decrease in the attenuation coefficient with the magnetic field for a scalar deformation potential was also predicted in Ref.~\cite{Pesin-Ilan:2020}. On the other hand, the mechanism of the sound dichroism uncovered in our work differs. Unlike Refs.~\cite{Sengupta-Garate:2020,Pesin-Ilan:2020}, it relies on the difference between the Fermi velocities and the intra-node relaxation times for the nodes of opposite chiralities. Furthermore, we show that the dichroism is also allowed by the strain-induced static pseudomagnetic field.

The paper is organized as follows. We introduce the model and the sound attenuation coefficient in Sec.~\ref{sec:Model}. Section~\ref{sec:CKT} is devoted to the chiral kinetic theory and the effect of propagating sound waves on the electron quasiparticle distribution function. The sound attenuation is analyzed and the numerical estimates are provided in Sec.~\ref{sec:Q}. The results are discussed and summarized in Secs.~\ref{sec:Discussions} and \ref{sec:Summary}, respectively. Technical details concerning the derivation of the energy dissipation rate and the collision integral are given in Appendixes~\ref{sec:app-0} and \ref{sec:app-1}, respectively. Throughout this paper, we use $k_{\rm B}=1$.

\section{Model and sound attenuation coefficient}
\label{sec:Model}

\subsection{Model of Weyl semimetal}
\label{sec:Model-0}

The effective low-energy Weyl Hamiltonian in the vicinity of the Weyl node $\alpha$ reads as
\begin{equation}
\label{Model-Weyl}
H_{\alpha} =\chi_{\alpha} v_{F,\alpha} \left(\mathbf{p}_{\alpha}\cdot\bm{\sigma}\right),
\end{equation}
where $\chi_{\alpha} =\pm$ is the chirality or, equivalently, the topological charge of the Weyl node, $v_{F,\alpha}$ is the Fermi velocity, $\mathbf{p}_{\alpha}$ is the momentum, and $\bm{\sigma}$ is the vector of the Pauli matrices acting in the pseudospin space. As we show in Secs.~\ref{sec:CKT-sol-B-b0} and \ref{sec:Q-B-b0},  difference between the Weyl nodes, e.g., different Fermi velocities, plays an important role in the sound attenuation in a magnetic field.

To realize a Weyl semimetal, the time-reversal (TR) and/or parity-inversion (PI) symmetries should be broken. In the case of the broken TR symmetry but preserved PI symmetry, the minimal model contains two Weyl nodes separated in momentum space by $2\mathbf{b}$. On the other hand, the minimal number of Weyl nodes for a TR-symmetric but PI-symmetry-broken model is 4. In the case where both symmetries are broken, the Weyl nodes in the minimal model could be separated in energy by $2b_0$ and in momentum by $2\mathbf{b}$. In material realizations of Weyl semimetals, the number of Weyl nodes could be significantly larger. For example, there are 24 Weyl nodes of two types in transition metal monopnictides (see, e.g., Ref.~\cite{Hasan-Huang:rev-2017}).

To describe the effects of weak dynamical strains, we introduce the deformation potential~\cite{Ziman:book}. For the sake of simplicity, we consider only its scalar part, i.e.,
\begin{equation}
\label{Model-Weyl-strain}
H_{\rm strain} =\lambda_{ij}^{(\alpha)}(\mathbf{p}_{\alpha})u_{ij}(t,\mathbf{r}).
\end{equation}
Here, $\lambda_{ij}^{(\alpha)}(\mathbf{p}_{\alpha})$ quantifies the strength of the scalar deformation potential for the node $\alpha$, $u_{ij}(t,\mathbf{r})=\left(\partial_i u_j +\partial_j u_i\right)/2$ is the strain tensor, and $\mathbf{u}(t,\mathbf{r})$ is the displacement vector that depends on time and coordinates. In this paper, we consider the case of plane sound waves
\begin{equation}
\label{CKT-sol-n-sound}
\mathbf{u}(t,\mathbf{r}) = \mathbf{u}_{0} e^{-i\omega t +i\mathbf{q}\cdot\mathbf{r}},
\end{equation}
where $\omega$ and $\mathbf{q}$ are the sound angular frequency and the wave vector, respectively. Henceforth, for simplicity, we suppress the arguments of the strain tensor. In general, $\lambda_{ij}^{(\alpha)}(\mathbf{p}_{\alpha})$ depends on momentum. This corresponds to the higher-order multipole modes of Fermi surface oscillations, which are the main source of the sound attenuation in usual single-valley metals~\cite{Abrikosov:book-1988}. In this paper, we focus on the momentum-independent deformation potential, i.e., we assume that $\lambda_{ij}^{(\alpha)}(\mathbf{p}_{\alpha})=\lambda_{ij}^{(\alpha)}$. This approximation is sufficient in order to capture the leading-order contribution to the sound absorption in multivalley systems.

It is well known~\cite{Abrikosov:book-1988} that, in the case of momentum-independent deformation potential, its valley-even component is strongly screened in conventional metals since it causes electric charge oscillations. On the other hand, the valley-odd component of the deformation potential in a multivalley conductor does not lead to electric charge oscillations, as it creates opposite in sign perturbations of electron distributions in the respective valleys. The electron scattering between the valleys results in the sound attenuation mechanism, similar to that of Debye relaxation losses in dielectrics~\cite{Debye:1912}; it remains effective even at perfect screening~\cite{Weinreich-White:1959,Gurevich-Efros:1963,Gantsevich-Gurevich:1967}. Electron transitions between the valleys are usually associated with a large (on the order of a Brillouin vector) momentum transfer. An external magnetic field does not considerably alter the corresponding matrix elements, as long as the magnetic length far exceeds the lattice parameter; this condition is satisfied by a large margin for any realistic field strength. The effect of magnetic field is qualitatively different in a Weyl semimetal, if the pattern of the odd-in-valley component of the deformation potential coincides with the pattern of nodes of opposite chirality. Magnetic field induces anomalous drift with opposite velocities for left- and right-handed electron quasiparticles. Due to the drifts, the perturbation of electron distribution caused by the valley-odd component of the deformation potential ceases to be charge neutral. That, in turn, activates the screening and suppresses the sound attenuation. We present a fuller qualitative explanation and interpretation of our results in Sec.~\ref{sec:Discussions}. As was noted in Ref.~\cite{Cortijo-Vozmediano:2016}, the chiral deformation potential can be induced in Weyl semimetals with broken PI symmetry where Weyl nodes are separated in energy. From the symmetry point of view, this part of the deformation potential requires coupling to phonon modes which are invariant under proper rotations (pseudoscalar phonons)~\cite{Rinkel-Garate:2017,Song-Dai:2016,Rinkel-Garate:2019}. Regardless of its origin, the chiral deformation potential corresponds to the antiphase motion of the Weyl nodes of opposite chiralities in energy space.

The energy spectrum in each Weyl node reads as
\begin{equation}
\label{Model-Weyl-eps-1}
\tilde{\epsilon}_{\alpha} = \epsilon_{\alpha} +\chi_{\alpha}b_0 +\lambda_{ij}^{(\alpha)}u_{ij},
\end{equation}
where $\epsilon_{\alpha}=\epsilon_{\alpha}(\mathbf{p}_{\alpha})$ is the energy dispersion in the absence of deformations at the node $\alpha$ and $b_0$ quantifies the separation between the Weyl nodes in energy space. For example, the Weyl nodes of opposite chiralities are located at different energies in SrSi$_2$~\cite{Huang-Hasan:2016,Singh-Bansil-SrSi2:2018}. In the case of the effective Hamiltonian given in Eq.~(\ref{Model-Weyl}), $\epsilon_{\alpha}=v_{F,\alpha}p_{\alpha}$. Finally, as in semiconductors~\cite{Gurevich-Efros:1963}, in the absence of electron transition between the valleys, each of the nodes establishes its own effective Fermi energy
\begin{equation}
\label{CKT-mu-1}
\mu_{\alpha} = \mu +\lambda_{ij}^{(\alpha)}u_{ij},
\end{equation}
where $\mu$ is the equilibrium Fermi energy measured from the Weyl nodes. This corresponds to the case where the local equilibrium charge density within each valley is conserved even in the presence of dynamical deformations. It is worth noting that such a configuration cannot exist in equilibrium for static deformations in real materials where even weak inter-node scattering processes eventually equalize the Fermi levels across all Weyl nodes.

\begin{figure}[!ht]
\begin{center}
\includegraphics[width=0.4\textwidth,clip]{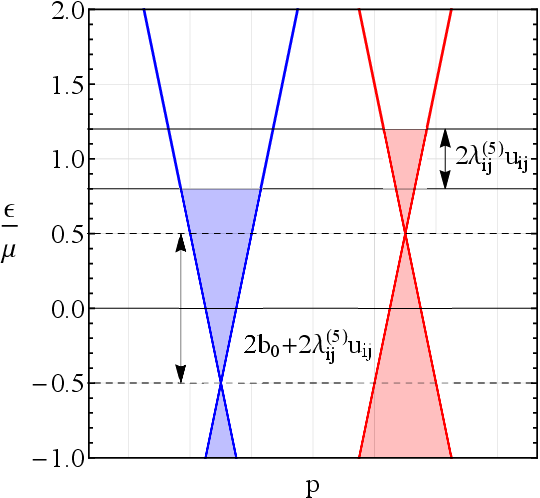}
\end{center}
\vspace{-0.5cm}
\caption{Schematic band structure of a Weyl semimetal with two nodes subject to the node-dependent or chiral deformation potential $\lambda_{ij}^{(\alpha)}u_{ij}=\chi_{\alpha}\lambda_{ij}^{(5)}u_{ij}$. In the absence of deformations, the Weyl nodes are separated by $2b_0$ in energy.
}
\label{fig:Model-schematic}
\end{figure}

A replica of the band structure displaying two representative Weyl nodes at nonzero chiral deformation potential $\lambda_{ij}^{(\alpha)}u_{ij}=\chi_{\alpha}\lambda_{ij}^{(5)}u_{ij}$ is sketched in Fig.~\ref{fig:Model-schematic}. As we demonstrate below, the energy shift induced by $\lambda_{ij}^{(5)}u_{ij}$ leads to a nontrivial scattering between Weyl nodes of opposite chiralities.

\subsection{Sound attenuation coefficient}
\label{sec:Q-gen}

In this section, we discuss the sound absorption in Weyl semimetals. The sound attenuation coefficient $\Gamma$ is defined as~\cite{Akhiezer-Liubarskii:1957,Gurevich-Pavlov:1971,Galperin-Kozub:rev-1979,Abrikosov:book-1988}
\begin{equation}
\label{Q-Gamma-def}
\Gamma = \frac{Q}{V I}.
\end{equation}
It quantifies the decay of the sound energy flux
\begin{equation}
\label{Q-I-def}
I = v_s\frac{\rho_{m} \langle |\partial_t \mathbf{u}|^2 \rangle_{T}}{2} = v_s\frac{\rho_{m} \omega^2 u_0^2}{2}
\end{equation}
within the sample. Here, $v_s$ is the sound velocity, $\rho_{m}$ is the mass density, $u_0$ is the magnitude of the displacement vector, $\langle \ldots \rangle_{T}$ denotes the time average, and $V$ is the system volume. Further, $Q$ is the dissipated energy per unit time. It can be conveniently expressed (see Appendix~\ref{sec:app-0}) in terms of the dynamical deformation potential $\lambda_{ij}^{(\alpha)} u_{ij}$ associated with the sound wave equation~(\ref{Model-Weyl-strain}) and the perturbed electron distribution evaluated to the first order in $u_{ij}$,
\begin{equation}
\label{Q-fin-fin}
\frac{Q}{V} = \sum_{\alpha}^{N_{W}} \frac{\nu_{\alpha}(\mu)}{2} \mbox{Re}\left\{i\omega\left(\lambda_{ij}^{(\alpha)} u_{ij}\right)^{*}\overline{n_{\alpha}}\right\}.
\end{equation}
Here, $N_{W}$ is the total number of Weyl nodes, and $\overline{n_{\alpha}}$ is the perturbed by the deformation potential electron distribution $n_{\alpha}(\mathbf{p}_{\alpha})\sim \lambda_{ij}^{(\alpha)}u_{ij}$ in the valley $\alpha$ averaged over the respective Fermi surface,
\begin{equation}
\label{CKT-equation-bar-def}
\overline{n_{\alpha}}= \frac{1}{\nu_{\alpha}(\mu)}\int \frac{d^3p_{\alpha}}{(2\pi \hbar)^3} \delta\left(\epsilon_{\alpha}+\chi_{\alpha}b_0-\mu\right) n_{\alpha}(\mathbf{p}_{\alpha}).
\end{equation}
We assume that temperature is small compared with the Fermi energy $\mu$ throughout this paper. The density of states (DOS) in the general equation~(\ref{Q-fin-fin}) for the specific model of Weyl nodes given in Eqs.~(\ref{Model-Weyl}) and (\ref{Model-Weyl-eps-1}) is
\begin{equation}
\label{CKT-equation-DOS}
\nu_{\alpha}(\mu) \equiv \int \frac{d^3p_{\alpha}}{(2\pi \hbar)^3}\delta\left(\epsilon_{\alpha}+\chi_{\alpha}b_0-\mu\right) = \frac{\left(\mu-\chi_{\alpha}b_0\right)^2}{2\pi^2 \hbar^3v_{F,\alpha}^3}.
\end{equation}
Since we focus on the case of classically weak magnetic fields $\omega_{c}\tau\ll1$ with $\omega_c$ being the cyclotron frequency and $\tau$ being the intra-node relaxation time, we neglect the dependence of the DOS on (pseudo)magnetic fields.

The full nonequilibrium distribution function for the quasiparticles from the Weyl node $\alpha$ reads
\begin{equation}
\label{CKT-equation-f-chi}
f_{\alpha}(\mathbf{p}_{\alpha}) =  f_{\alpha}^{(0)}(\mathbf{p}_{\alpha})-(\partial_{\epsilon_{\alpha}}f_{\alpha}^{(0)}) n_{\alpha}(\mathbf{p}_{\alpha}),
\end{equation}
where $f_{\alpha}^{(0)}(\mathbf{p}_{\alpha})$ is the Fermi-Dirac distribution function, which describes electron quasiparticles in local equilibrium, and
\begin{equation}
\label{CKT-equation-df-chi}
(\partial_{\epsilon_{\alpha}}f_{\alpha}^{(0)}) \approx -\delta\left(\epsilon_{\alpha}+\chi_{\alpha}b_0-\mu\right)
\end{equation}
for temperatures low compared with the Fermi energy.

In the case of equal densities of states $\nu_{\alpha}(\mu)$ in all Weyl nodes, $\nu_{\alpha}(\mu)=\nu(\mu)$, and the deformation potential $\lambda_{ij}^{(\alpha)}=\lambda_{ij}+\chi_{\alpha}\lambda_{ij}^{(5)}$ uniform across the nodes of same chirality, Eq.~(\ref{Q-fin-fin}) simplifies to
\begin{widetext}
\begin{equation}
\label{Q-fin-fin-equal}
\frac{Q}{V} = -\frac{\nu(\mu) \omega}{2}\mbox{Im}\left\{\left(\lambda_{ij} u_{ij}\right)^{*}\overline{n} +\left(\lambda_{ij}^{(5)} u_{ij}\right)^{*}\overline{n_{5}}\right\}
\quad{\rm at} \quad \nu_{\alpha}(\mu)=\nu(\mu), \quad \lambda_{ij}^{(\alpha)}=\lambda_{ij}+\chi_{\alpha}\lambda_{ij}^{(5)}.
\end{equation}
\end{widetext}
Here, the averaged distribution functions $\overline{n}=\sum_{\alpha}^{N_{W}}\overline{n_{\alpha}}$ and $\overline{n_{5}}=\sum_{\alpha}^{N_{W}}\chi_{\alpha}\overline{n_{\alpha}}$ are linear combinations of the perturbed electron distributions in valleys $\alpha$ averaged over the respective Fermi surfaces. The local charge density is suppressed in the limit of strong screening, $\overline{n}=0$, and the first term in the curly brackets in Eq.~(\ref{Q-fin-fin-equal}) vanishes.

Finally, we notice that the energy dissipation rate in Eq.~(\ref{Q-fin-fin}) agrees with that in Ref.~\cite{Galperin-Kozub:rev-1979}. We would like to emphasize that using Eq.~(\ref{Q-fin-fin}) is equivalent to the conventional way of evaluation of $Q$ via entropy production~\cite{Akhiezer-Liubarskii:1957,Landau:t10-1995}. However, we find the form of Eq.~(\ref{Q-fin-fin}) better suited for our purposes, as it allows us to see the effect of symmetries between the valleys on the energy dissipation and spares us from a separate evaluation of the Joule and inter-valley contributions to $Q$. As we show in Sec.~\ref{sec:Q-B}, our results for the sound absorption in a magnetic field agree (in the appropriate limits) with the recent findings in Ref.~\cite{Pesin-Ilan:2020}, where the entropy production was calculated.

For our qualitative estimates, we ignore the anisotropy of the deformation potential in the attenuation coefficient and estimate it as $\lambda_{ij}^{(\alpha)} u_{ij}\sim i\lambda^{(\alpha)} u_0 q$. Then, the final expression for the sound attenuation coefficient in Eq.~(\ref{Q-Gamma-def}) reads as
\begin{equation}
\label{Q-Gamma-fin}
\Gamma = \frac{q}{v_s\rho_{m} \omega u_0} \sum_{\alpha}^{N_{W}} \nu_{\alpha}(\mu)\, \mbox{Re}\left\{\lambda^{(\alpha)} \overline{n_{\alpha}}\right\}.
\end{equation}
Thus, the sound attenuation is determined by the DOS $\nu_{\alpha}(\mu)$, the deformation potential $\propto \lambda^{(\alpha)}$, and the averaged nonequilibrium part of the distribution function $\overline{n_{\alpha}}$. Under the simplifying assumptions regarding the DOS in Weyl nodes and the chiral deformation potential used in Eq.~(\ref{Q-fin-fin-equal}), the attenuation coefficient reduces to:
\begin{eqnarray}
\label{Q-Gamma-fin-equal}
&&\Gamma = \frac{\nu(\mu) q}{v_s\rho_{m} \omega u_0} \mbox{Re}\left\{\lambda \overline{n} +\lambda^{(5)}\overline{n_{5}}\right\}\quad{\rm at} \quad \nu_{\alpha}(\mu)=\nu(\mu), \nonumber\\ &&\lambda_{ij}^{(\alpha)}=\lambda_{ij}+\chi_{\alpha}\lambda_{ij}^{(5)}.
\end{eqnarray}
To describe the deviations from the equilibrium caused by dynamical deformations and determine $\overline{n_{\alpha}}$, we employ the chiral kinetic theory (CKT) in the next section.

\section{Chiral kinetic theory}
\label{sec:CKT}

\subsection{General equations of chiral kinetic theory}
\label{sec:CKT-equations}

The semiclassical kinetic equation for Weyl quasiparticles reads as~\cite{Son:2013,Stephanov:2012,Son-Spivak:2013}
\begin{widetext}
\begin{eqnarray}
&&\partial_t f_{\alpha} +\frac{1}{\Theta_{\alpha}}
\left\{\Big(-e\tilde{\mathbf{E}}_{\alpha} -\frac{e}{c}\left[\mathbf{v}_{\alpha}\times \mathbf{B}_{\alpha}\right] +\frac{e^2}{c}(\tilde{\mathbf{E}}_{\alpha}\cdot\mathbf{B}_{\alpha})\mathbf{\Omega}_{\alpha}\Big)\cdot\partial_{\mathbf{p}_{\alpha}} f_{\alpha}
+\Big(\mathbf{v}_{\alpha} -e\left[\tilde{\mathbf{E}}_{\alpha}\times\mathbf{\Omega}_{\alpha}\right]
-\frac{e}{c}(\mathbf{v}_{\alpha}\cdot\mathbf{\Omega}_{\alpha})\mathbf{B}_{\alpha}\Big)\cdot \bm{\nabla} f_{\alpha}\right\} \nonumber\\
&&=I_{\rm intra}\left[f_{\alpha}\right] + I_{\rm inter}\left[f_{\alpha}\right].
\label{CKT-equation-kinetic-equation}
\end{eqnarray}
\end{widetext}
Here, $f_{\alpha}$ is the distribution function for electron quasiparticles at the node $\alpha$ defined in Eq.~(\ref{CKT-equation-f-chi}), $\mathbf{v}_{\alpha}=\partial_{\mathbf{p}_{\alpha}}\tilde{\epsilon}_{\alpha}$ is the quasiparticle velocity, $-e\tilde{\mathbf{E}}_{\alpha}$ is the force acting on an electron ($e>0$),
\begin{equation}
\label{CKT-equations-E-chi-def}
e\tilde{\mathbf{E}}_{\alpha} =e\mathbf{E}+\bm{\nabla}\tilde{\epsilon}_{\alpha}\,,
\end{equation}
the energy spectrum $\tilde{\epsilon}_{\alpha}$ is defined in Eq.~(\ref{Model-Weyl-eps-1}) and depends on spatial coordinates via the deformation potential $\lambda_{ij}^{(\alpha)}u_{ij}$, and vector $\mathbf{B}_{\alpha}$ combines the external magnetic and pseudomagnetic fields; see Eq.~(\ref{CKT-equations-B-chi-def}) for an explicit expression for $\mathbf{B}_{\alpha}$ in a symmetric case. The Berry curvature $\mathbf{\Omega}_{\alpha}=\mathbf{\Omega}_{\alpha}(\mathbf{p}_\alpha)$ for the Hamiltonian of Eq.~(\ref{Model-Weyl}) takes the form
\begin{equation}
\mathbf{\Omega}_{\alpha}({\mathbf p}_{\alpha}) =\chi_{\alpha} \hbar\frac{\mathbf{p}_{\alpha}}{2p_{\alpha}^3}
\label{CKT-equation-Berry-monopole}
\end{equation}
for electrons with energies above the Weyl point, and $\Theta_{\alpha}=\left[1-e \left(\mathbf{B}_{\alpha}\cdot \mathbf{\Omega}_{\alpha}\right)/c\right]$ is the renormalization of the phase-space volume. The sound attenuation is determined by the Fermi surface properties, so, in the following, we will encounter only $\mathbf{\Omega}_{\alpha}({\mathbf p}_{F,\alpha})$ where ${\mathbf p}_{F,\alpha}$ parametrizes points on the Fermi spheres around the respective Weyl nodes $\alpha$. In the case of a linear energy spectrum discussed in Sec.~\ref{sec:Model-0}, $p_{F,\alpha}=(\mu-\chi_{\alpha}b_0)/v_{F,\alpha}$. The collision integral on the right-hand side in Eq.~(\ref{CKT-equation-kinetic-equation}) is a sum of the intra- and inter-node terms $I_{\rm intra}\left[f_{\alpha}\right]$ and $I_{\rm inter}\left[f_{\alpha}\right]$, respectively. Their explicit form in the $\tau$ approximation is given in Appendix~\ref{sec:app-1}; see Eqs.~(\ref{app-1-intra-2}) and (\ref{app-1-Icoll-inter-fin}).

The kinetic equation (\ref{CKT-equation-kinetic-equation}) should also be amended with Maxwell's equations. Since the sound and Fermi velocities are much smaller than the speed of light, we use a quasistatic approximation in which dynamical magnetic and solenoidal electric fields are neglected. Then, only the Gauss law should be taken into account. It reads as
\begin{equation}
\label{CKT-equation-Gauss}
\bm{\nabla}\cdot\mathbf{E} = -4\pi e \sum_{\alpha}^{N_{W}} \nu_{\alpha}(\mu)\overline{n_{\alpha}}
\end{equation}
and is determined by the nonequilibrium part of the distribution function [see also Eq.~(\ref{CKT-equation-f-chi})].
The combination of the kinetic equation (\ref{CKT-equation-kinetic-equation}) and the Gauss law (\ref{CKT-equation-Gauss}) comprise the full formulation of the self-consistent problem for the response of a Weyl semimetal to acoustic waves propagating in it.

As is shown in Sec.~\ref{sec:Q-gen}, the energy dissipation rate and the sound attenuation coefficient are determined by the averaged over the Fermi surface nonequilibrium part of the distribution function $\overline{n_{\alpha}}$. To calculate this function, we linearize Eq.~(\ref{CKT-equation-kinetic-equation}) in weak strains $u_{ij}$ and use the ansatz in Eq.~(\ref{CKT-equation-f-chi}). Then, Eq.~(\ref{CKT-equation-kinetic-equation}) reads
\begin{widetext}
\begin{eqnarray}
&&\left\{\partial_t +\left[\mathbf{v}_{\alpha}-\frac{e}{c}\left(\mathbf{v}_{\alpha}\cdot\mathbf{\Omega}_{\alpha}({\mathbf p}_{F,\alpha})\right)\mathbf{B}_{\alpha}\right]\cdot\bm{\nabla} \right\}n({\mathbf p}_{F,\alpha})\nonumber\\
&&+\left[e\left(\mathbf{v}_{\alpha}\cdot\mathbf{E}\right) -\frac{e^2}{c} \left(\mathbf{v}_{\alpha}\cdot\mathbf{\Omega}_{\alpha}({\mathbf p}_{F,\alpha})\right) \left(\mathbf{E}\cdot\mathbf{B}_{\alpha}\right) +\sum_{\beta}^{N_{W}} \frac{n({\mathbf p}_{F,\alpha})-\overline{n_{\beta}}}{\tau_{\alpha,\beta}}\right] \nonumber\\
&&= -\left[\lambda_{ij}^{(\alpha)} \left(\mathbf{v}_{\alpha}\cdot \bm{\nabla}u_{ij} \right) -\lambda_{ij}^{(\alpha)}\frac{e}{c}\left(\mathbf{v}_{\alpha}\cdot\mathbf{\Omega}_{\alpha}({\mathbf p}_{F,\alpha})\right) \left(\mathbf{B}_{\alpha}\cdot\bm{\nabla}u_{ij} \right) +u_{ij}\sum_{\beta}^{N_{W}} \frac{\lambda_{ij}^{(\alpha)}-\lambda_{ij}^{(\beta)}}{\tau_{\alpha,\beta}}\right].
\label{CKT-equation-kinetic-equation-lin}
\end{eqnarray}
\end{widetext}
Here, $n({\mathbf p}_{F,\alpha})$ is the nonequilibrium part of the distribution function at the Fermi level. We used the dispersion relation of Eq.~(\ref{Model-Weyl-eps-1}) and the explicit form of the collision integrals in the $\tau$ approximation given in Eqs.~(\ref{app-1-intra-2}) and (\ref{app-1-Icoll-inter-fin-aver}) in Appendix~\ref{sec:app-1}. The relaxation rates
\begin{equation}
\label{CKT-equation-tau-alpha-beta-def}
\frac{1}{\tau_{\alpha,\beta}} = \frac{2\pi}{\hbar} |A_{\alpha,\beta}|^2 \nu_{\beta}(\mu)
\end{equation}
are determined by the intra- and inter-node scattering amplitudes $A_{\alpha,\beta}$. In the following, we abbreviate the notation for the intra-node relaxation time: $\tau_{\alpha,\alpha}\equiv\tau_{\alpha}$. In addition, in the derivation of Eq.~(\ref{CKT-equation-kinetic-equation-lin}), we neglected the phase-space volume renormalization $\Theta_{\alpha}=\left[1-e \left(\mathbf{B}_{\alpha}\cdot \mathbf{\Omega}_{\alpha}\right)/c\right]$ and the contribution of the magnetic moment to the energy dispersion. Accounting for it is equivalent~\cite{Xiao-Niu:rev-2010,Son:2013} to the replacement $\epsilon_{\alpha}\to\epsilon_{\alpha}\left[1+ e \left(\bm{\Omega}_{\alpha}\cdot\mathbf{B}_{\alpha}\right)/c\right]$. As is estimated at the end of Sec.~\ref{sec:Q-B}, these terms lead only to a small contribution to the sound attenuation coefficient compared with the effect of the chiral anomaly.

\subsection{Solutions to the kinetic equation}
\label{sec:CKT-sol}

In this section, we consider the case of weak inter-node ($\alpha\neq\beta$) scattering, $\tau_{\alpha}\ll\tau_{\alpha,\beta}$, and $\tau_{\alpha} \omega\ll 1$. We checked that the above assumptions hold well for realistic numerical values presented in Sec.~\ref{sec:Q-B}. The short intra-node relaxation time allows us to retain only the first two harmonics in the expansion of the nonequilibrium part of the distribution function:
\begin{equation}
\label{CKT-equation-n-expand}
n_{\alpha}(\mathbf{p}_{\alpha}) \approx n_{\alpha}^{(0)} + n_{\alpha}^{(1)} \cos{\theta_{\alpha}},
\end{equation}
where $\theta_{\alpha}$ is the angle between $\mathbf{v}_{\alpha}$ and $\bm{\nabla}$, and $n_{\alpha}^{(0)}\approx \overline{n_{\alpha}}$. By following the standard procedure, we use Eq.~(\ref{CKT-equation-kinetic-equation-lin}), separate the contributions with different powers of $\cos{\theta_{\alpha}}$, and solve for $n_{\alpha}^{(1)}$. Then, by using the obtained solution, we calculate the current density
\begin{equation}
\label{CKT-equation-node-j}
\mathbf{j}_{\alpha} = \mathbf{j}_{\alpha}^{\rm (CME)} +\mathbf{j}_{\alpha}^{\rm (diff)}.
\end{equation}
In the above equation, we separated the chiral (pseudo)magnetic effect (CME) $\mathbf{j}_{\alpha}^{\rm (CME)}$~\cite{Vilenkin:1980,Fukushima:2008} and diffusion intra-node $\mathbf{j}_{\alpha}^{\rm (diff)}$ currents. They read as
\begin{widetext}
\begin{eqnarray}
\label{CKT-equation-node-j-CME}
\mathbf{j}_{\alpha}^{\rm (CME)} &=& \frac{e^2}{c} \int \frac{d^3p_{\alpha}}{(2\pi \hbar)^3} \left(\mathbf{v}_{\alpha}\cdot\mathbf{\Omega}_{\alpha}({\mathbf p}_{F,\alpha})\right)\mathbf{B}_{\alpha} n_{\alpha}^{(0)}\delta\left(\epsilon_{\alpha}+\chi_{\alpha}b_0-\mu\right) = \chi_{\alpha} \frac{e^2}{4\pi^2\hbar^2 c} \mathbf{B}_{\alpha} \overline{n_{\alpha}},\\
\label{CKT-equation-node-j-intra}
\mathbf{j}_{\alpha}^{(\rm diff)} &=&  -e\int \frac{d^3p_{\alpha}}{(2\pi \hbar)^3} \mathbf{v}_{\alpha} n_{\alpha}^{(1)}\cos{\theta_{\alpha}} \delta\left(\epsilon_{\alpha}+\chi_{\alpha}b_0-\mu\right) =e\nu_{\alpha}(\mu)D_{\alpha} \left[\bm{\nabla}\overline{n_{\alpha}} +e\mathbf{E}+\lambda_{ij}^{(\alpha)}\bm{\nabla}u_{ij}\right],
\end{eqnarray}
\end{widetext}
where
\begin{equation}
\label{CKT-equation-tD-def}
D_{\alpha}=\frac{v_{F,\alpha}^2 \tau_{\alpha}}{3}
\end{equation}
is the diffusion coefficient. Notice that the term $\left[\tilde{\mathbf{E}}_{\alpha}\times\mathbf{\Omega}_{\alpha}({\mathbf p}_{F,\alpha})\right]$ present in electron velocity [see Eq.~(\ref{CKT-equation-kinetic-equation})] does not contribute to the current density in the linear order in $u_{ij}$.

By averaging Eq.~(\ref{CKT-equation-kinetic-equation-lin}) over the Fermi surface and using Eq.~(\ref{CKT-equation-node-j}), we derive the following kinetic equation:
\begin{widetext}
\begin{equation}
\label{CKT-equation-n-chi}
\partial_t \overline{n_{\alpha}} -\frac{1}{e\nu_{\alpha}(\mu)}\left(\bm{\nabla}\cdot\mathbf{j}_{\alpha}\right) -\frac{e^2}{c}\left(\left[\mathbf{E}+\ \frac{1}{e}\lambda_{ij}^{(\alpha)}\bm{\nabla}u_{ij}\right]\cdot\mathbf{B}_{\alpha}\right) \overline{\left(\mathbf{\Omega}_{\alpha}\cdot\mathbf{v}_{\alpha}\right)}
= -\sum_{\beta}^{N_{W}}\frac{\overline{n_{\alpha}} -\overline{n_{\beta}}}{\tau_{\alpha,\beta}} 
-u_{ij} \sum_{\beta}^{N_{W}} \frac{\lambda_{ij}^{(\alpha)}-\lambda_{ij}^{(\beta)}}{\tau_{\alpha,\beta}},
\end{equation}
\end{widetext}
which together with Eqs.~(\ref{CKT-equation-node-j})--(\ref{CKT-equation-node-j-intra}) as well as the Gauss law (\ref{CKT-equation-Gauss}) defines the response $\overline{n_{\alpha}}$ to the dynamic strain $u_{ij}$.

The first two terms on the left-hand side of Eq.~(\ref{CKT-equation-n-chi}) correspond to the conventional continuity equation. The chiral anomaly is described by the third term, where
\begin{equation}
\label{CKT-equation-Omega-v-bar}
\overline{\left(\mathbf{\Omega}_{\alpha}\cdot\mathbf{v}_{\alpha}\right)} =\left|\mathbf{\Omega}_{\alpha}({\mathbf p}_{F,\alpha})\right| v_{F,\alpha}
= \frac{\chi_{\alpha}}{4\pi^2 \hbar^2 \nu_{\alpha}(\mu)};
\end{equation}
cf. Eq.~(\ref{CKT-equation-Berry-monopole}). The collision integral on the right-hand side contains the term describing usual inter-node scattering (the first term). The second term in the collision integral originates from the different effective Fermi energies of the Weyl nodes of opposite chiralities (see also Fig.~\ref{fig:Model-schematic} for the schematic band structure). It is worth noting that this term and the diffusion current were not accounted for in Ref.~\cite{Spivak:2016}.

In the case of sound-induced dynamical strains, the deformation potential and the nonequilibrium part of the distribution function have a plane-wave dependence on time and coordinates; see Eq.~(\ref{CKT-sol-n-sound}).
Therefore, by using the explicit form of the currents given in Eqs.~(\ref{CKT-equation-node-j-CME}) and (\ref{CKT-equation-node-j-intra}), we rewrite Eq.~(\ref{CKT-equation-n-chi}) as
\begin{widetext}
\begin{eqnarray}
\label{CKT-sol-n-chi}
&&\left[ q^2D_{\alpha} -i\left(\mathbf{v}_{\Omega,\alpha}\cdot\mathbf{q}\right) -i\omega\right] n_{\alpha} -ieD_{\alpha}\left(\mathbf{q}\cdot\mathbf{E}\right) -\left(\mathbf{v}_{\Omega,\alpha}\cdot\mathbf{E}\right) +\sum_{\beta}^{N_{W}}\frac{n_{\alpha} -n_{\beta}}{\tau_{\alpha,\beta}} \nonumber\\
&& = -u_{ij} \left[\lambda_{ij}^{(\alpha)} q^2D_{\alpha} -i\lambda_{ij}^{(\alpha)}  \left(\mathbf{v}_{\Omega,\alpha}\cdot\mathbf{q}\right) +\sum_{\beta}^{N_{W}}\frac{\lambda_{ij}^{(\alpha)}-\lambda_{ij}^{(\beta)}}{\tau_{\alpha,\beta}}\right].
\end{eqnarray}
\end{widetext}
Here, to simplify the notations, we omitted the overbar in the averaged distribution function, i.e., $\overline{n_{\alpha}}\to n_{\alpha}$, and introduced the anomalous velocity
\begin{equation}
\label{CKT-sol-vOmega-def}
\mathbf{v}_{\Omega,\alpha}= \chi_{\alpha} |\mathbf{\Omega}_{\alpha}({\mathbf p}_{F,\alpha})| \frac{v_{F,\alpha}}{c} e\mathbf{B}_{\alpha} = \chi_{\alpha} \frac{e}{4\pi^2 c\hbar^2 \nu_{\alpha}(\mu)} \mathbf{B}_{\alpha}.
\end{equation}
In an external magnetic field, $\mathbf{B}_{\alpha} =\mathbf{B}$, it corresponds to the drift of electrons with opposite chiralities in opposite directions.

As one can see from Eq.~(\ref{CKT-sol-n-chi}), diffusion [see the second term in Eq.~(\ref{CKT-sol-n-chi})] and the chiral anomaly [see the third term in Eq.~(\ref{CKT-sol-n-chi})] introduce the dependence on the electric field $\mathbf{E}$. The Gauss law (\ref{CKT-equation-Gauss}) allows one to express $\mathbf{E}$ as a linear combination of $n_{\alpha}$, thus completing the formulation of the linear response problem. We will provide its explicit solution under simplifying assumptions.

Let us consider Weyl semimetals where there is a symmetry between the pairs ($\alpha,-\alpha$) of Weyl nodes. Specifically, we assume that the deformation potential can be parametrized as
\begin{equation}
\label{CKT-sol-lambda-symm}
\lambda_{ij}^{(\alpha)} = \lambda_{ij} +\chi_{\alpha} \lambda_{ij}^{(5)},
\end{equation}
the two relaxation times
\begin{equation}
\label{CKT-sol-tau5-def}
\frac{1}{\tau_{5,\pm\alpha}} = \sum_{\beta}^{N_{W}} \frac{|\chi_{\pm\alpha}-\chi_{\beta}|}{\tau_{\pm\alpha,\beta}}
\end{equation}
are the same for all pairs, and
\begin{equation}
\label{CKT-equations-B-chi-def}
\mathbf{B}_{\alpha} =\mathbf{B}+\chi_{\alpha}\mathbf{B}_{5}.
\end{equation}
Here the pseudomagnetic field $\mathbf{B}_{5}$ can be generated by static strains such as bending or torsion~\cite{Zhou-Shi:2013,Cortijo-Vozmediano:2015,Grushin:2016,Pikulin-Franz:2016} (see also Ref.~\cite{Ilan-Pikulin:rev-2019} for a review). As an example of the corresponding system with such a symmetry, we mention Weyl semimetals where the pairs of Weyl nodes are well separated in momentum space. Then, only the scattering rates between the nodes $\alpha$ and $-\alpha$ of opposite chiralities inside each pair $\alpha$ contribute to $\tau_{5,\alpha}$, i.e., $\tau_{5,\alpha}\approx \tau_{\alpha,-\alpha}$. In the presence of said symmetry, the matrix of the linear system in Eq.~(\ref{CKT-sol-n-chi}) becomes block diagonal, and one can consider kinetic equations only for a single pair of Weyl nodes with chiralities $\chi_{-\alpha}=-\chi_{\alpha}$.

To shorten the notations, we introduce the square of the inverse length~\cite{AM:book-1976}
\begin{equation}
q_{\rm TF}^2=4\pi e^2\sum_{\alpha}^{N_{W}}\nu_{\alpha}(\mu)
\label{CKT-sol-qTF}
\end{equation}
of the Thomas-Fermi screening in Eq.~(\ref{CKT-sol-n-chi}). By expressing the electric field in terms of $n_{\alpha}$ with the help of the Gauss law (\ref{CKT-equation-Gauss}) and using Eq.~(\ref{CKT-sol-qTF}), we rewrite the kinetic equations for a single pair $\alpha=\pm 1$ of Weyl nodes as
\begin{widetext}
\begin{eqnarray}
\label{CKT-sol-n}
&&\left[\frac{1}{\tau_{5,\alpha}} +q^2D_{\alpha} -i\left(\mathbf{v}_{\Omega,\alpha}\cdot \mathbf{q}\right)  -i\omega\right] n_{\alpha} +\left[q^2D_{\alpha} -i \left(\mathbf{v}_{\Omega,\alpha}\cdot \mathbf{q}\right)\right] \frac{q_{\rm TF}^2}{q^2} \frac{\nu_{\alpha}(\mu)n_{\alpha}+\nu_{-\alpha}(\mu)n_{-\alpha}}{\nu_{\alpha}(\mu)+\nu_{-\alpha}(\mu)} \nonumber\\
&&- \frac{n_{\alpha}+n_{-\alpha}}{2\tau_{5,\alpha}}
= -\lambda_{ij}u_{ij} \left[q^2D_{\alpha} -i\left(\mathbf{v}_{\Omega,\alpha}\cdot \mathbf{q}\right)\right] -\chi_{\alpha}\lambda_{ij}^{(5)}u_{ij} \left[\frac{1}{\tau_{5,\alpha}} +q^2D_{\alpha} -i\left(\mathbf{v}_{\Omega,\alpha}\cdot \mathbf{q}\right)\right].
\end{eqnarray}
\end{widetext}

In what follows, we solve Eq.~(\ref{CKT-sol-n}) for symmetric Weyl nodes ($b_0=0$, $v_{F,\alpha}=v_F$, $\tau_{\alpha}=\tau$, and $\tau_{5,\alpha}=\tau_5$) in two cases: (i) $B\neq0$ and $B_5=0$ as well as (ii) $B=0$ and $B_5\neq0$. In addition, the solutions for nonsymmetric nodes ($b_0\neq0$, $v_{F,\alpha}\neq v_{F,-\alpha}$, $\tau_{\alpha}\neq\tau_{-\alpha}$, and $\tau_{5,\alpha}\neq\tau_{5,-\alpha}$) at $B\neq0$ and $B_5=0$ are also considered.

\subsubsection{Solution of the kinetic equations at a finite magnetic field}
\label{sec:CKT-sol-B}

In the case of symmetric Weyl nodes and $B_5=0$, one can simplify Eq.~(\ref{CKT-sol-n}) to
\begin{widetext}
\begin{eqnarray}
\label{CKT-sol-B-n-2}
&&\left[\frac{1}{\tau_{q}}-i\chi_{\alpha} \left(\mathbf{v}_{\Omega}\cdot \mathbf{q}\right)  -i\omega\right] n_{\alpha} -\left[\frac{1}{\tau_{5}}  -q_{\rm TF}^2 D+ i\chi_{\alpha}\left(\mathbf{v}_{\Omega}\cdot \mathbf{q}\right) \frac{q_{\rm TF}^2}{q^2}\right] \frac{\left(n_{\alpha}+n_{-\alpha}\right)}{2} \nonumber\\
&&= -\lambda_{ij}u_{ij} \left[q^2D -i\chi_{\alpha} \left(\mathbf{v}_{\Omega}\cdot \mathbf{q}\right)\right] -\chi_{\alpha} \lambda_{ij}^{(5)}u_{ij} \left[\frac{1}{\tau_{q}} -i\chi_{\alpha} \left(\mathbf{v}_{\Omega}\cdot \mathbf{q}\right)\right],
\end{eqnarray}
\end{widetext}
where $\mathbf{v}_{\Omega}=\chi_{\alpha}\mathbf{v}_{\Omega,\alpha}$ and the effective scattering rate
\begin{equation}
\label{CKT-sol-B-tauq-1}
\frac{1}{\tau_{q}} = \frac{1}{\tau_{5}} +q^2 D
\end{equation}
includes both intra- and inter-node scattering processes. Notice that $\mathbf{v}_{\Omega}\propto \mathbf{B}$ is a TR-odd axial vector. The full solution to Eq.~(\ref{CKT-sol-B-n-2}) reads
\begin{widetext}
\begin{eqnarray}
\label{CKT-sol-B-n-sol}
n_{\alpha} &=& -\frac{\lambda_{ij}u_{ij}}{M} \left[ q^2 D \left(\frac{1}{\tau_{q}} -i\omega\right)  -\chi_{\alpha} \omega \left(\mathbf{v}_{\Omega}\cdot \mathbf{q}\right) +\left(\mathbf{v}_{\Omega}\cdot \mathbf{q}\right)^2\right] \nonumber\\
&-&\chi_{\alpha} \frac{\lambda_{ij}^{(5)}u_{ij}}{M} \left[ \frac{\left(q_{\rm TF}^2+q^2\right)D -i\omega}{\tau_{q}} -\chi_{\alpha} \omega \left(\mathbf{v}_{\Omega}\cdot \mathbf{q}\right) +(q_{\rm TF}^2+q^2)\left(\mathbf{v}_{\Omega}\cdot \hat{\mathbf{q}}\right)^2\right],
\end{eqnarray}
\end{widetext}
where $\hat{\mathbf{q}}=\mathbf{q}/q$ and
\begin{eqnarray}
\label{CKT-sol-B-D-sol}
M &=& \left[\left(q_{\rm TF}^2+q^2\right)D-i\omega\right]\left(1/\tau_{q} -i\omega\right) \nonumber\\
&+& (q_{\rm TF}^2+q^2)\left(\mathbf{v}_{\Omega}\cdot \hat{\mathbf{q}}\right)^2
\end{eqnarray}
is proportional to the determinant of the system in Eq.~(\ref{CKT-sol-B-n-2}).

In material realizations of Weyl semimetals, the Fermi energy $\mu$ is small compared with that in metals. For example, $\mu=10-30~\mbox{meV}$ in a typical Weyl semimetal TaAs~\cite{Huang-Chen-TaAs:2015,Arnold-Felser:2016b}. Still the free-carrier density is high enough to result in a fairly short Thomas-Fermi screening length, so that the conditions
\begin{equation}
\label{CKT-sol-B-simpl}
q_{\rm TF}\gg q, \quad \sigma\gg\omega, \quad \sigma\gg\tau_{5}\omega^2
\end{equation}
are easily satisfied. Here $\sigma=q_{\rm TF}^2D/(4\pi)$ is the electric conductivity at $B=B_5=0$. The first of the conditions (\ref{CKT-sol-B-simpl}) allows one to disregard the effect of the valley-even part of the deformation potential $\lambda_{ij}$ in Eq.~(\ref{CKT-sol-B-n-sol}): In agreement with expectations~\cite{Weinreich-White:1959} (see also Refs.~\cite{Gurevich-Efros:1963,Galperin-Kozub:rev-1979}), its contribution to $n_{\alpha}$ scales as $(q/q_{\rm TF})^2$. Using Eq.~(\ref{CKT-sol-B-simpl}), we simplify the general solution in Eqs.~(\ref{CKT-sol-B-n-sol}) and (\ref{CKT-sol-B-D-sol}), and find $n_{\alpha}$ as
\begin{eqnarray}
\label{CKT-sol-B-n-sol-expl}
n_{\alpha} &=& -\chi_{\alpha} \lambda_{ij}^{(5)}u_{ij} \frac{1/\tau_{q} +\left(\mathbf{v}_{\Omega}\cdot \hat{\mathbf{q}}\right)^2/D}{ 1/\tau_{q}+ \left(\mathbf{v}_{\Omega}\cdot \hat{\mathbf{q}}\right)^2/D-i\omega}, \nonumber\\
n &=& \sum_{\alpha}^{N_{W}}n_{\alpha}=0, \,\,\,  n_{5} = \sum_{\alpha}^{N_{W}}\chi_{\alpha}n_{\alpha}=N_W \chi_{\alpha}n_\alpha.
\end{eqnarray}
As we showed in Sec.~\ref{sec:Q-gen}, since the attenuation coefficient is determined by the real part of $n_5$ [see Eq.~(\ref{Q-Gamma-fin-equal})] it is already evident that the combination of the electrostatic screening and the intra-node momentum relaxation has a profound effect on how the magnetic field affects the sound dissipation. We discuss this in detail in Sec.~\ref{sec:Q-B}.

\subsubsection{Solution of the kinetic equation at a finite pseudomagnetic field}
\label{sec:CKT-sol-B5}

Let us now consider the case in which the system is subject only to a pseudomagnetic field $\mathbf{B}_5$. The kinetic equation~(\ref{CKT-sol-n}) at $B=0$ and symmetric Weyl nodes reads as
\begin{widetext}
\begin{eqnarray}
\label{CKT-sol-B5-n}
&&\left[\frac{1}{\tau_{q}} -i\left(\mathbf{v}_{\Omega,5}\cdot \mathbf{q}\right) -i\omega\right] n_{\alpha} 
- \left[\frac{1}{\tau_{5}} -q_{\rm TF}^2 D +  i\left(\mathbf{v}_{\Omega,5}\cdot \mathbf{q}\right) \frac{q_{\rm TF}^2}{q^2} \right] \frac{\left(n_{\alpha}+n_{-\alpha}\right)}{2} \nonumber\\
&&= -\lambda_{ij}u_{ij} \left[q^2D -i\left(\mathbf{v}_{\Omega,5}\cdot \mathbf{q}\right) \right] -\chi_{\alpha}\lambda_{ij}^{(5)}u_{ij} \left[\frac{1}{\tau_{q}} -i\left(\mathbf{v}_{\Omega,5}\cdot \mathbf{q}\right)\right],
\end{eqnarray}
\end{widetext}
where $\mathbf{v}_{\Omega,5}=\mathbf{v}_{\Omega,\alpha}$ is a TR-odd polar vector. Indeed, $\mathbf{v}_{\Omega,5}\propto \mathbf{B}_5$, where $\mathbf{B}_5$ breaks both TR and PI symmetries. Under the conditions of Eq.~(\ref{CKT-sol-B-simpl}), its solution is
\begin{eqnarray}
\label{CKT-sol-B5-n-sol-fin}
n_{\alpha} &=& -\chi_{\alpha} \lambda_{ij}^{(5)}u_{ij} \frac{1/\tau_{q} -i \left(\mathbf{v}_{\Omega,5}\cdot \mathbf{q}\right)}{1/\tau_{q} -i\left(\mathbf{v}_{\Omega,5}\cdot \mathbf{q}\right)-i\omega}, \nonumber\\
n &=& \sum_{\alpha}^{N_{W}}n_{\alpha}=0, \quad n_5=\sum_{\alpha}^{N_{W}}\chi_{\alpha}n_{\alpha}.
\end{eqnarray}
Notice that the direction of the pseudomagnetic field depends on the pattern of Weyl node pairs. Therefore the summation over all Weyl nodes should be performed with care. We discuss a few corresponding examples in Sec.~\ref{sec:Q-B5}.

By comparing Eqs.~(\ref{CKT-sol-B-n-sol-expl}) and (\ref{CKT-sol-B5-n-sol-fin}), it is evident that the magnetic and pseudomagnetic fields affect the distribution functions differently. In particular, the distribution function $n_{\alpha}$ in Eq.~(\ref{CKT-sol-B5-n-sol-fin}) depends on the direction of the pseudomagnetic field $\mathbf{B}_5$. This observation is not surprising because $\mathbf{B}_5$, which can be induced by static strains such as torsion or bending, breaks both the symmetry between the Weyl nodes and crystal symmetries. As we show in Sec.~\ref{sec:Q-B5}, the pseudomagnetic field has a profound effect on the sound attenuation in Weyl semimetals leading to distinct attenuation coefficients for the sound propagating along and opposite to the pseudomagnetic field.

\subsubsection{Solution of the kinetic equations at a finite magnetic field: nonsymmetric Weyl nodes}
\label{sec:CKT-sol-B-b0}

Finally, we analyze the case of nonsymmetric Weyl nodes, i.e., $b_0\neq0$, $v_{F,\alpha}\neq v_{F,-\alpha}$, $\tau_{\alpha}\neq\tau_{-\alpha}$, and $\tau_{5,\alpha}\neq\tau_{5,-\alpha}$. The solution to Eq.~(\ref{CKT-sol-n}) reads
\begin{widetext}
\begin{eqnarray}
\label{CKT-sol-B-b0-n}
n_{\alpha} &=& -\chi_{\alpha}\lambda^{(5)}_{ij}u_{ij}\frac{2\nu_{-\alpha}(\mu)}{\nu_{\alpha}(\mu)+\nu_{-\alpha}(\mu)} \frac{1}{\tilde{M}}\Bigg\{\tau_{5,\alpha} \left[q^2D_{-\alpha} -i\left(\mathbf{v}_{\Omega,-\alpha}\cdot\mathbf{q}\right)\right]+\tau_{5, -\alpha} \left[q^2D_{\alpha}-i\left(\mathbf{v}_{\Omega,\alpha}\cdot\mathbf{q}\right) \right] \nonumber\\
&+& 2\tau_{5,\alpha}\tau_{5,-\alpha} \left[q^2D_{\alpha}-i\left(\mathbf{v}_{\Omega,\alpha}\cdot\mathbf{q}\right) \right]\left[q^2D_{-\alpha} -i\left(\mathbf{v}_{\Omega,-\alpha}\cdot\mathbf{q}\right)\right]\Bigg\},
\end{eqnarray}
\end{widetext}
with
\begin{widetext}
\begin{eqnarray}
\label{CKT-sol-B-b0-N}
\tilde{M} &=& \tau_{5,\alpha} \left[q^2D_{-\alpha} -i\left(\mathbf{v}_{\Omega,-\alpha}\cdot\mathbf{q}\right) \right] +\tau_{5,-\alpha} \left[q^2D_{\alpha} -i\left(\mathbf{v}_{\Omega,\alpha}\cdot\mathbf{q}\right) \right] \nonumber\\
&+&\frac{2\tau_{5,\alpha}\tau_{5,-\alpha}}{\nu_{\alpha}(\mu) +\nu_{-\alpha}(\mu)} \Bigg\{\nu_{\alpha}(\mu)\left[q^2D_{\alpha} -i\left(\mathbf{v}_{\Omega,\alpha}\cdot\mathbf{q}\right)\right] \left[q^2D_{-\alpha} -i \left(\mathbf{v}_{\Omega,-\alpha}\cdot\mathbf{q}\right)-i\omega\right] \nonumber\\
&+& \nu_{-\alpha}(\mu)\left[q^2D_{\alpha}-i \left(\mathbf{v}_{\Omega,\alpha}\cdot\mathbf{q}\right)-i\omega\right]\left[q^2D_{-\alpha} -i\left(\mathbf{v}_{\Omega,-\alpha}\cdot\mathbf{q}\right)\right] \Bigg\}.
\end{eqnarray}
\end{widetext}
In deriving Eqs.~(\ref{CKT-sol-B-b0-n}) and (\ref{CKT-sol-B-b0-N}), we used Eq.~(\ref{CKT-sol-B-simpl}) and retained only the terms surviving~\footnote{Notice that, as in the case of symmetric Weyl nodes considered in Sec.~\ref{sec:CKT-sol-B}, the valley-even part of the deformation potential, $\lambda_{ij}u_{ij}$, does not contribute to the averaged distribution function in the leading order in $q/q_{\rm TF}$} in the limit $1/q_{\rm TF}\to 0$. The solution remains cumbersome even at $1/q_{\rm TF}\to 0$. Therefore, in the following, we consider small deviations from the symmetry:
\begin{eqnarray}
\label{CKT-sol-B-b0-tD-exp-vF}
D_{\alpha} &=& D +\chi_{\alpha} \delta D = D\left[1- \chi_{\alpha}\frac{\delta \nu(\mu)}{\nu(\mu)}\right] +\chi_{\alpha} \delta \tilde{D}, \nonumber\\
\nu_{\alpha}(\mu) &=& \nu(\mu) +\chi_{\alpha} \delta \nu(\mu).
\end{eqnarray}
Here, $D=(D_{\alpha}+D_{-\alpha})/2$, while the difference $\delta D=\chi_{\alpha}(D_{\alpha}-D_{-\alpha})/2$ is assumed to be small, $\delta D\ll D$. Notice also that while $D$ is a true scalar, $\delta D$ is a pseudoscalar. Similar definitions and assumptions are made for other variables.

In particular, $\delta \tau_{5}$ and $\delta \mathbf{v}_{\Omega}$ depend only on the DOS deviations $\delta \nu(\mu)$ similarly to the second term in the square brackets in Eq.~(\ref{CKT-sol-B-b0-tD-exp-vF}). [We assume also the case of well-separated Weyl node pairs, as discussed after Eq.~(\ref{CKT-equations-B-chi-def}).] It is important that $\delta D$ contains an additional term $\delta \tilde{D}$, which is determined by different Fermi velocities and/or intra-node scattering amplitudes in the nodes $\alpha$ and $-\alpha$. As we show below, this term is responsible for the odd-in-magnetic field term in $n_{\alpha}$.

Expanding the solution (\ref{CKT-sol-B-b0-n}) up to the first order in $\delta \nu(\mu)$ and $\delta \tilde{D}$, $n_{\alpha}= n_{\alpha}^{(0)} +n_{\alpha}^{(1)}+\dots$, we obtain
\begin{widetext}
\begin{eqnarray}
\label{CKT-sol-B-b0-n-exp-1-vF-2}
n_{\alpha}^{(1)} &=& -\chi_{\alpha} n_{\alpha}^{(0)}\frac{\delta \nu(\mu)}{\nu(\mu)} -2\chi_{\alpha} \lambda_{ij}^{(5)}u_{ij} \frac{ \left(\mathbf{v}_{\Omega}\cdot\mathbf{q}\right)\omega}{\left[1/\tau_{q} +\left(\mathbf{v}_{\Omega}\cdot \hat{\mathbf{q}}\right)^2/D -i\omega\right]^2} \frac{\delta \tilde{D}}{D}, \\
\label{CKT-sol-B-b0-n-exp-5-vF}
n_{5}^{(1)} &=& \sum_{\alpha}^{N_{W}}\chi_{\alpha}n_{\alpha}^{(1)} = -2N_{W}\lambda_{ij}^{(5)}u_{ij}\frac{ \left(\mathbf{v}_{\Omega}\cdot\mathbf{q}\right)\omega}{\left[1/\tau_{q} +\left(\mathbf{v}_{\Omega}\cdot \hat{\mathbf{q}}\right)^2/D-i\omega\right]^2} \frac{\delta \tilde{D}}{D},
\end{eqnarray}
\end{widetext}
where the zeroth-order solution $n_{\alpha}^{(0)}$ is given by Eq.~(\ref{CKT-sol-B-n-sol-expl}) found in the limit of symmetric Weyl nodes. While the zeroth-order solution is even in $\mathbf{B}\cdot\hat{\mathbf{q}}$, odd powers of $\mathbf{B}\cdot\hat{\mathbf{q}}$ appear in the first-order correction given in Eq.~(\ref{CKT-sol-B-b0-n-exp-1-vF-2}) at $\delta\tilde D\neq 0$. The corresponding component survives the summation over the nodes, see Eq.~(\ref{CKT-sol-B-b0-n-exp-5-vF}). It is easy to check that, as expected for the strong screening conditions (\ref{CKT-sol-B-simpl}), deviations of the resulting electric charge vanish, i.e., $\sum_{\alpha}^{N_{W}}\nu_{\alpha}(\mu)n_{\alpha}^{(0)}+\sum_{\alpha}^{N_{W}}\nu(\mu)n_{\alpha}^{(1)}=0$.

In the next section, by using the obtained solutions, we discuss the sound attenuation in Weyl semimetals.

\section{Sound attenuation}
\label{sec:Q}

In parallel with Sec.~\ref{sec:CKT-sol}, we evaluate the sound attenuation in three limits. We start with the results for symmetric Weyl nodes and nonzero magnetic field (while $B_5=0$), then consider the effect of the pseudomagnetic field (at $B=0$), and conclude this Section with the results specified for nonsymmetric Weyl nodes at $B_5=0$.

\subsection{Sound attenuation in a magnetic field}
\label{sec:Q-B}

By using Eqs.~(\ref{Q-Gamma-fin-equal}) and (\ref{CKT-sol-B-n-sol-expl}), we find the following attenuation coefficient for the nonzero magnetic field $\mathbf{B}$ and symmetric Weyl nodes:
\begin{equation}
\label{Q-B-Gamma-B}
\Gamma (\mathbf{B},\omega, \hat{\mathbf{q}}) = \frac{N_{W}\nu(\mu) |\lambda^{(5)}|^2}{v_s \rho_{\rm m}} \frac{q^2 \left[1/\tau_{q} +\left(\mathbf{v}_{\Omega}\cdot \hat{\mathbf{q}}\right)^2/D\right]}{\left[1/\tau_{q} +\left(\mathbf{v}_{\Omega}\cdot \hat{\mathbf{q}}\right)^2/D\right]^2 +\omega^2},
\end{equation}
where $q=\omega/v_s$. In the absence of magnetic fields, $B=0$ (i.e., $v_{\Omega}=0$), the sound attenuation coefficient reads as $\Gamma (\omega)= \left[N_{W}\nu(\mu)|\lambda^{(5)}|^2/(v_s \rho_{\rm m})\right] \left\{\tau_{q}q^2/\left[1 +\left(\tau_{q}\omega\right)^2\right]\right\}$. It agrees with the results obtained in Refs.~\cite{Weinreich-White:1959,Gurevich-Efros:1963,Gantsevich-Gurevich:1967} for multivalley semiconductors and in Ref.~\cite{Pesin-Ilan:2020} for Weyl semimetals. At $\tau_q\omega\ll1$, it scales as $\omega^2$. The same scaling is valid also for single-valley conductors with momentum-dependent deformation potential~\cite{Akhiezer-Liubarskii:1957,Abrikosov:book-1988}.

As one can see from Eq.~(\ref{Q-B-Gamma-B}), the sound attenuation coefficient monotonically decreases with the increase in magnetic field $\mathbf{B}$. By using the typical parameter values [see Eq.~(\ref{Q-num-TaAs}) below] it is easy to check that the condition $\tau_{q}\omega \ll1$ holds well for a wide range of intra- and inter-node relaxation times. This allows us to simplify Eq.~(\ref{Q-B-Gamma-B}),
\begin{equation}
\label{Q-B-Gamma-B_limit}
\Gamma (\mathbf{B},\omega, \hat{\mathbf{q}}) = \frac{N_{W}\nu(\mu) |\lambda^{(5)}|^2}{v_s \rho_{\rm m}} \frac{q^2}{1/\tau_{q} +\left(\mathbf{v}_{\Omega}\cdot \hat{\mathbf{q}}\right)^2/D},
\end{equation}
where the dependence of $\mathbf{v}_{\Omega}$ on $\mathbf{B}$ is given by Eq.~(\ref{CKT-sol-vOmega-def}). Further expansion of Eq.~(\ref{Q-B-Gamma-B_limit}) in $\mathbf{B}$ would reproduce the result of Ref.~\cite{Pesin-Ilan:2020}.

We used semiclassical kinetic equations to derive Eqs.~(\ref{Q-B-Gamma-B}) and (\ref{Q-B-Gamma-B_limit}). Therefore we confine our consideration to non-quantizing magnetic fields. Furthermore, we did not account for the effect of magnetic field on the intra-valley electron dynamics. At an arbitrary angle between $\mathbf{B}$ and $\hat{\mathbf{q}}$, this limits our consideration to classically weak magnetic fields; cf. Ref.~\cite{Pesin-Ilan:2020}. The latter constraint is eased for the field $\mathbf{B}$ aligned with $\hat{\mathbf{q}}$, as we are considering spherical Fermi surfaces. In that respect, there is a similarity between the evaluation of $\Gamma(B)$ and the evaluation of the conductivity tensor component along the direction of a non-quantizing magnetic field~\cite{Lifshitz-Kaganov:1957}: In a metal with a spherical Fermi surface, cyclotron motion of electrons does not affect these two quantities.

\begin{figure}[!ht]
\begin{center}
\includegraphics[width=0.4\textwidth,clip]{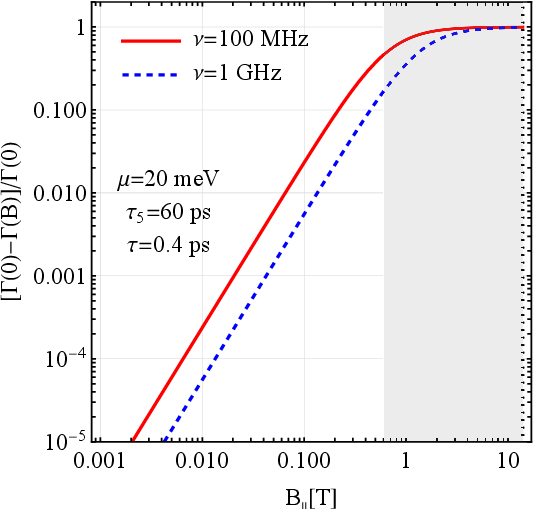}
\end{center}
\vspace{-0.5cm}
\caption{
The magnetic-field dependent part of the sound attenuation coefficient for the field $B_{\parallel}$ applied along the sound propagation direction. For the parameters given in Eq.~(\ref{Q-num-TaAs}), one may use Eq.~(\ref{Q-B-Gamma-B_limit}) at frequencies $\nu\equiv\omega/(2\pi)\lesssim 100~\mbox{MHz}$ (red solid line). For illustration, we also plot the attenuation coefficient given in Eq.~(\ref{Q-B-Gamma-B}) for a much higher frequency, $\nu=1~\mbox{GHz}$ (blue dashed line), at which $q^2D \tau_5>1$. The gray shaded area corresponds to a classically strong magnetic field $\omega_{c}\tau\gg1$. Here, $\omega_c =eB_{\alpha} v_{F, \alpha}^2/\left[c(\mu-\chi_{\alpha} b_0)\right]$ is the cyclotron frequency for the node $\alpha$. The crossover to a quantizing magnetic field region with $\hbar\omega_c/(\sqrt{2}\mu)\gtrsim1$ is indicated by the black dotted line.
}
\label{fig:Q-B-num-Gamma}
\end{figure}

We present the dependence of the normalized magnetic-field dependent part of the sound attenuation coefficient in Fig.~\ref{fig:Q-B-num-Gamma}. In plotting it, we used Eqs.~(\ref{Q-B-Gamma-B}) and (\ref{Q-B-Gamma-B_limit}) as well as parameters
\begin{eqnarray}
\label{Q-num-TaAs}
&&v_{\rm F}\approx 3\times 10^7~\mbox{cm/s}, \,\,\,  \mu\approx 20~\mbox{meV}, \,\,\, \tau_{5}\approx 6\times10^{-11}~\mbox{s}, \nonumber\\
&&\tau\approx 3.8\times 10^{-13}\mbox{s}, \,\,\, v_s \approx 2.8\times 10^5~\mbox{cm/s},
\end{eqnarray}
quoted~\cite{Arnold-Felser:2016b,Zhang-Hasan-TaAs:2016,Laliberte-Quilliam:2019} for the Weyl semimetal TaAs. As one can see from Fig.~\ref{fig:Q-B-num-Gamma}, the suppression of the sound attenuation coefficient by magnetic field is significant for experimentally feasible frequencies of ultrasound and for $B$ approaching the regime of classically strong magnetic fields.

Our Eqs.~(\ref{Q-B-Gamma-B}) and (\ref{Q-B-Gamma-B_limit}) differ substantially from the sound attenuation predictions of Ref.~\cite{Spivak:2016}. There are two major differences distinguishing the current work from the simplifications accepted in Ref.~\cite{Spivak:2016}, as we account for (i) a small but finite intra-node relaxation time and (ii) electrostatic screening of the electric fields accompanying sound waves in a Weyl semimetal subject to a magnetic field. It is the combination of these effects that leads to the discrepancy. To demonstrate this explicitly, let us consider the model case of a very fast intra-node relaxation, $D\to0$. Then, the attenuation coefficient due to the valley-dependent part of the deformation potential reads
\begin{eqnarray}
\label{Q-B-Gamma-tD0}
&&\lim_{D\to0}\Gamma(\mathbf{B},\omega, \hat{\mathbf{q}}) = \frac{N_{W}\nu(\mu)|\lambda^{(5)}|^2}{v_s \rho_{\rm m}} \nonumber\\ 
&&\times\frac{\tau_{5} q^2}{(\tau_5\omega)^2\left[1-\left(q_{\rm TF}^2+q^2\right)\left(\mathbf{v}_{\Omega}\cdot \hat{\mathbf{q}}\right)^2/\omega^2\right]^2 +1},
\end{eqnarray}
where we used Eqs.~(\ref{CKT-sol-B-n-sol}) and (\ref{CKT-sol-B-D-sol}) without any assumption about the ratio $q/q_{\rm TF}$. The magnetic field dependence first appears here in the second order in the magnetic field:
\begin{eqnarray}
\label{Q-B-deltaGamma-tD0}
&&\lim_{D\to0} \Gamma(\mathbf{B},\omega, \hat{\mathbf{q}}) \!=\! \lim_{D\to0}\Gamma(\omega)+\delta\Gamma(\mathbf{B},\omega, \hat{\mathbf{q}}), \nonumber\\
&&\delta\Gamma(\mathbf{B},\omega, \hat{\mathbf{q}}) \!=\! \frac{2N_{W}\nu(\mu)|\lambda^{(5)}|^2}{v_s\rho_{\rm m}}\frac{(\tau_{5}\omega)^2}{v_s^2} \tau_{5} \nonumber\\
&&\times \left(q_{\rm TF}^2+q^2\right) \left(\mathbf{v}_{\Omega}\cdot \hat{\mathbf{q}}\right)^2.
\end{eqnarray}
To obtain the final form of $\delta\Gamma(\mathbf{B},\omega, \hat{\mathbf{q}})$ in the above equation, we considered the limit of low frequency, $\tau_{5}\omega\ll 1$. Without screening, $q_{\rm TF}=0$, we find a qualitative~\footnote{Notice that Ref.~\cite{Spivak:2016} misses a factor $\sim(\tau_{5}\omega)^2$, which is present in Eq.~(\ref{Q-B-deltaGamma-tD0}). This discrepancy comes from an overly simplified form of the collision integral used in Ref.~\cite{Spivak:2016}. In particular, the last term in the collision integral (\ref{app-1-Icoll-inter-1-fin}) is absent in that study} agreement with the conclusions of Ref.~\cite{Spivak:2016}: magnetic field would enhance sound attenuation. Furthermore, as follows from Eq.~(\ref{Q-B-deltaGamma-tD0}), screening ($q_{\rm TF}\neq0$) strengthens the chiral anomaly effect. Thus, the qualitative difference between the results in Eqs.~(\ref{Q-B-Gamma-B}) and (\ref{Q-B-Gamma-tD0}) stems from the intra-node diffusion at finite $D$, along with the presence of electric fields accompanying non-uniform chiral currents. We provide a more detailed discussion of the interplay of the chiral anomaly, screening, and diffusion in Sec.~\ref{sec:Discussions}. Regretfully, even in a strongly-disordered material with the mean free path of the order of the Fermi wavelength, the diffusion coefficient is high enough to invalidate Eq.~(\ref{Q-B-Gamma-tD0}). By the same token, the associated with the chiral anomaly propagating collective mode~\cite{Spivak:2016} turns into an overdamped one due to diffusion. One can see this by examining the roots of the polynomial $M(\omega)$ in Eq.~(\ref{CKT-sol-B-D-sol}).

Finally, to justify our approximation in Eq.~(\ref{CKT-equation-kinetic-equation-lin}), let us estimate the effects related to the magnetic moment and the renormalization of the phase-space volume $\Theta_{\alpha}$. To start with, we notice that there is a profound qualitative difference between the effects of the chiral anomaly on one side and the effects of the phase-space renormalization and the orbital magnetic moment on the other. Indeed, the phase-space volume renormalization provides a correction to the density of states~\cite{Xiao-Niu:rev-2010,Son:2013} [see also Eq.~(\ref{app-1-cc-DOS})] which modifies the distributions in each of the nodes separately. On the other hand, the chiral anomaly leads to the relative drift of the distributions at nodes of opposite chiralities. To quantify the comparison between the two types of the magnetic field effects, we turn to Eq.~(\ref{Q-B-Gamma-B_limit}). According to it, the role of the chiral anomaly in the sound absorption is governed by the parameter $\tau_q v_{\Omega}^2/D$. On the other hand, the effect of the magnetic field due to the renormalization of the phase-space volume and the magnetic moment is controlled by the parameter $v_{\Omega}^2/v_F^2$ [see, e.g., the definition of $\Theta_{\alpha}$ after Eq.~(\ref{CKT-equation-Berry-monopole})]~\footnote{Notice that due to symmetry reasons, odd-in-magnetic field terms may not appear in the sound attenuation coefficient}. Comparing the respective parameters, we see that the chiral anomaly dominates in the magnetic field dependence of the sound attenuation for $\tau/\tau_5 +(v_F \tau q)^2\ll 1$. Since the inter-node scattering time $\tau_5$ is much larger than the intra-node one $\tau$ and the mean-free path $v_F \tau \lesssim 1/q$ for a low-frequency sound, the effects of the phase-space renormalization and the magnetic moment could be indeed neglected compared with the chiral anomaly contribution in the sound attenuation. We provide a detailed qualitative discussion of the chiral anomaly effect on the sound attenuation in Sec.~\ref{sec:Discussions}.

\subsection{Sound attenuation in a pseudomagnetic field}
\label{sec:Q-B5}

In this section, we consider the sound attenuation in the pseudomagnetic field $\mathbf{B}_5$. As in Sec.~\ref{sec:CKT-sol-B5}, we assume that the Weyl nodes are symmetric. Furthermore, since the pseudomagnetic field depends on the configuration of the Weyl nodes, we focus on a system with a single pair of nodes. The extension to the case of multiple Weyl nodes is discussed at the end of this section. By using Eqs.~(\ref{Q-Gamma-fin-equal}) and (\ref{CKT-sol-B5-n-sol-fin}), we derive
\begin{equation}
\label{Q-B5-Gamma-All}
\Gamma(\mathbf{B}_5,\omega, \hat{\mathbf{q}}) =
\frac{2\nu(\mu) |\lambda^{(5)}|^2}{v_s \rho_{\rm m}} \frac{q^2/\tau_{q}}{1/\tau_{q}^2+\left[\omega +\left(\mathbf{v}_{\Omega,5}\cdot\mathbf{q}\right)\right]^2}.
\end{equation}
Here, $(\mathbf{v}_{\Omega,5}\cdot\mathbf{q})$ is a true scalar because $\mathbf{v}_{\Omega,5}\propto\mathbf{B}_5$ is a polar vector. Among the most notable features of $\Gamma(\mathbf{B}_5,\omega, \hat{\mathbf{q}})$ is its dependence on the direction of the pseudomagnetic field with respect to the sound wave vector, see also Eq.~(\ref{CKT-sol-vOmega-def}) for the dependence of $\mathbf{v}_{\Omega,5}$ on $\mathbf{B}_5$. Indeed, depending on the sign of  $\left(\mathbf{v}_{\Omega,5}\cdot\mathbf{q}\right)$, the pseudomagnetic field can either reduce or enhance the sound absorption. Furthermore, $\Gamma(\mathbf{B}_5,\omega, \hat{\mathbf{q}})$ is a nonmonotonic function of $\mathbf{B}_5$ which has a maximum, $\Gamma_{\rm max}(\omega, \hat{\mathbf{q}}) = 2\nu(\mu) |\lambda^{(5)}|^2 \tau_{q}\omega^2/(v_s^3 \rho_{\rm m})$, at $\omega=-\left(\mathbf{v}_{\Omega,5}\cdot\mathbf{q}\right)$.

One can easily extract the odd-in-pseudomagnetic-field part of the attenuation coefficient. The result at $\tau_{q}\left|\omega +\left(\mathbf{v}_{\Omega,5}\cdot\mathbf{q}\right)\right|\ll1$ reads
\begin{eqnarray}
\label{Q-B5-Gamma-All-weak-B5-odd}
&&\Gamma(\mathbf{B}_5,\omega, \hat{\mathbf{q}})-\Gamma(\mathbf{B}_5,\omega, -\hat{\mathbf{q}}) \nonumber\\
&&\approx
-\frac{8\nu(\mu) \left|\lambda^{(5)}\right|^2}{v_s^3 \rho_{\rm m}} (\tau_{q}\omega)^3 \left(\mathbf{v}_{\Omega,5}\cdot\mathbf{q}\right).
\end{eqnarray}
The above equation clearly shows that the sound attenuation is different when the sound propagates along or opposite to the pseudomagnetic field. Therefore we dub this effect the \emph{pseudomagnetic sound dichroism}.

\begin{figure*}[!ht]
\centering
\subfigure[]{\includegraphics[height=0.4\textwidth]{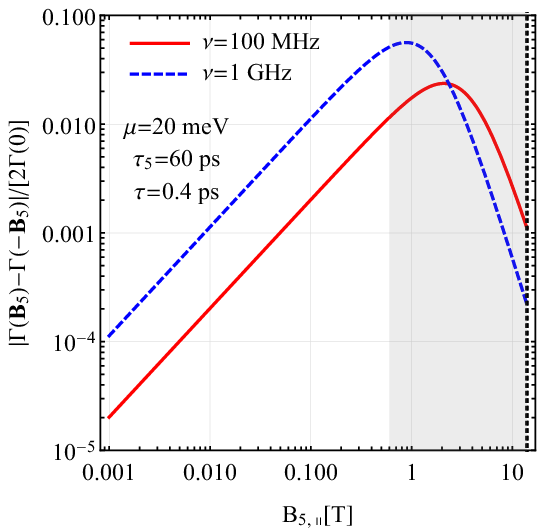}}
\hspace{0.1\textwidth}
\subfigure[]{\includegraphics[height=0.4\textwidth]{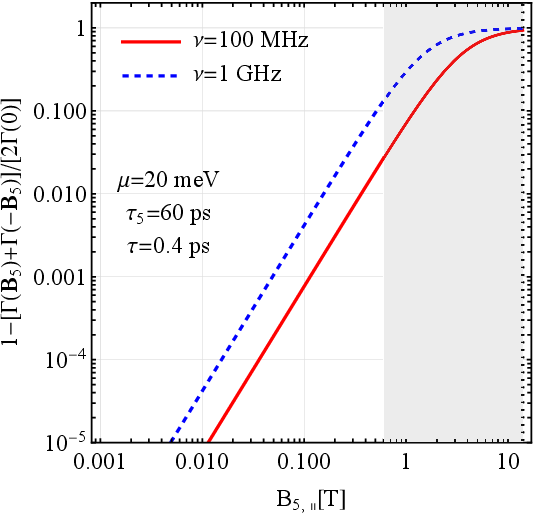}}
\vspace{-0.5cm}
\caption{
Dependence of the (a) odd- and (b) even-in-pseudomagnetic field parts of the attenuation coefficient on the pseudomagnetic field applied along the sound propagation direction $B_{5,\parallel}$. We used the following sound frequencies $\nu\equiv\omega/(2\pi)$: $\nu=100~\mbox{MHz}$ (red solid line) and $\nu=1~\mbox{GHz}$ (blue dashed line), which correspond to the regimes $q^2D \tau_5<1$ and $q^2D \tau_5>1$, respectively. In addition, we used the parameters of Eq.~(\ref{Q-num-TaAs}). The gray shaded area corresponds to a classically strong pseudomagnetic field. The crossover to a quantizing pseudomagnetic field region is indicated by the black dotted line.
}
\label{fig:Q-B5-num-Gamma-close}
\end{figure*}

To estimate the effects of the pseudomagnetic field on the sound absorption, we show the odd- and even-in-pseudomagnetic field parts of the attenuation coefficient in Figs.~\ref{fig:Q-B5-num-Gamma-close}(a) and \ref{fig:Q-B5-num-Gamma-close}(b), respectively. Our numerical estimates suggest that the dichroism is weak and reaches $\sim 4~\%$ for a field $B_5\sim 1~\mbox{T}$. The dependence of the even-in-pseudomagnetic field part is qualitatively similar to that for $B$; cf. Figs.~\ref{fig:Q-B-num-Gamma} and \ref{fig:Q-B5-num-Gamma-close}(b). Indeed, the pseudomagnetic field generally reduces the sound attenuation coefficient. The reduction is strong for large fields comparable to classically strong $\mathbf{B}_5$. It is worth noting that attainable values of strain-induced pseudomagnetic fields are estimated to be about $0.3~\mbox{T}$ in a twisted wire~\cite{Pikulin-Franz:2016} and about $15~\mbox{T}$ for a bent film~\cite{Pikulin-Franz:2017}.

Finally, let us comment on the sound absorption in a system with multiple Weyl nodes. Unlike the case with a nonzero magnetic field considered in Sec.~\ref{sec:Q-B}, the pseudomagnetic field itself depends on the configuration of Weyl nodes. Therefore, in general, one needs to use Eqs.~(\ref{Q-Gamma-fin}) and (\ref{CKT-sol-n-chi}) to obtain the sound attenuation coefficient. However, as we discussed after Eq.~(\ref{CKT-sol-vOmega-def}), certain symmetries can be used to identify the pairs of Weyl nodes and the corresponding pseudomagnetic fields. For example, in the simplest model of TR symmetric Weyl semimetals, there are two pairs of Weyl nodes such that the pseudomagnetic fields in each pair have opposite directions. Therefore, the sound attenuation coefficient is given by the sum of the attenuation coefficients obtained in Eq.~(\ref{Q-B5-Gamma-All}) with the opposite signs of the field $\mathbf{B}_5$. No pseudomagnetic sound dichroism appears in this case. The dichroism could be observed if the TR symmetry is broken, e.g., in an intrinsically magnetic material such as EuCd$_2$As$_2$~\cite{Soh-Boothroyd:2019,Ma-Shi:2019}.

\subsection{Sound attenuation dichroism in a magnetic field}
\label{sec:Q-B-b0}

In this section, we calculate the sound attenuation in Weyl semimetals with nonsymmetric Weyl nodes subject to a magnetic field. As in Sec.~\ref{sec:CKT-sol-B-b0}, we focus on the stemming from the difference between the Weyl nodes first-order correction $\Gamma^{(1)}(\mathbf{B},\omega, \hat{\mathbf{q}})$ to the attenuation coefficient. In particular, we assume that the differences between the densities of states $\left[\nu_{\alpha}(\mu)-\nu_{-\alpha}(\mu)\right]/\nu(\mu)\ll1$, the Fermi velocities $(v_{F,\alpha}-v_{F,-\alpha})/v_F\ll1$, and the diffusion constants $(D_{\alpha}-D_{-\alpha})/D\ll1$ are small. This is sufficient to capture the odd-in-magnetic-field part of $\Gamma$. We use the sound attenuation coefficient in Eq.~(\ref{Q-Gamma-fin}) and the solution given in Eq.~(\ref{CKT-sol-B-b0-n-exp-5-vF}). The resulting first-order correction is
\begin{eqnarray}
\label{Q-B-b0-Gamma-1-2}
&&\!\Gamma^{(1)}(\mathbf{B},\omega, \hat{\mathbf{q}}) = \frac{2N_{W} |\lambda^{(5)}|^2 \nu(\mu) \omega}{v_s \rho_{m}} \nonumber\\
&&\!\times \frac{q^2 \left[1/\tau_{q} +\left(\mathbf{v}_{\Omega}\cdot\hat{\mathbf{q}}\right)^2/D \right]}{\left\{\! \left[1/\tau_{q} +\left(\mathbf{v}_{\Omega}\cdot\hat{\mathbf{q}}\right)^2/D\right]^2 +\omega^2\! \right\}^2} \frac{\delta \tilde{D}}{D} \! \left(\mathbf{v}_{\Omega}\cdot\mathbf{q}\right),
\end{eqnarray}
where $\delta \tilde{D}$ is determined by the difference
of the Fermi velocities and/or the intra-node scattering amplitudes around the Weyl nodes $\alpha$ and $-\alpha$. Because $\delta \tilde{D}$ and $\left(\mathbf{v}_{\Omega}\cdot\mathbf{q}\right)$ are pseudoscalars, the resulting attenuation coefficient is a scalar. As one can see from Eq.~(\ref{Q-B-b0-Gamma-1-2}), there is a dependence of the sound absorption on the relative orientation of the magnetic field and the sound wave vector. Indeed, the sign of the coefficient $\Gamma^{(1)}(\mathbf{B},\omega, \hat{\mathbf{q}})$ is flipped together with the direction of the magnetic field with respect to the sound wave vector, resulting in the \emph{magnetic sound dichroism}. A similar effect was predicted in Refs.~\cite{Sengupta-Garate:2020,Pesin-Ilan:2020}, whose origin, however, relies on the difference of the deformation potentials between different nodes. In the limit $\tau_{q}\omega\ll1$ and for small magnetic fields, the scaling with $\omega$ and $\tau_{5}$ agrees with that of Ref.~\cite{Pesin-Ilan:2020}.

It is interesting to compare the odd-in-magnetic-field attenuation coefficient with its pseudomagnetic counterpart. Upon expanding Eq.~(\ref{Q-B-b0-Gamma-1-2}) in $\tau_{q}\omega \ll1$ and weak magnetic fields $(\hat{\mathbf{q}}\cdot\mathbf{v}_{\Omega})^2\ll D/\tau_{q}$, its comparison with Eq.~(\ref{Q-B5-Gamma-All-weak-B5-odd}) yields
\begin{equation}
\label{Q-B-b0-Gamma-1-compare}
\frac{\Gamma^{(1)}(\mathbf{B},\omega, \hat{\mathbf{q}})-\Gamma^{(1)}(\mathbf{B},\omega, -\hat{\mathbf{q}})}{\Gamma(\mathbf{B}_5,\omega, \hat{\mathbf{q}})-\Gamma(\mathbf{B}_5,\omega, -\hat{\mathbf{q}})} =  -\frac{\left(\mathbf{v}_{\Omega}\cdot\mathbf{q}\right)}{\left(\mathbf{v}_{\Omega,5}\cdot\mathbf{q}\right)} \frac{\delta \tilde{D}}{D}.
\end{equation}
Therefore, the magnetic sound dichroism is expected to be weaker than its pseudomagnetic counterpart at equal respective fields.

\section{Discussion}
\label{sec:Discussions}

The effect of magnetic or pseudo-magnetic field on the sound attenuation in Weyl semimetals uncovered in this paper has a simple explanation. To understand this and clarify the crucial role of screening and intra-node relaxation, let us recall the well established in the 1950s and 1960s theory of sound attenuation in metals, semimetals, and doped semiconductors. In a ``single-valley" metal, a momentum-independent deformation potential does not produce sound attenuation~\cite{Abrikosov:book-1988}. Indeed, under strong screening conditions, the deformation potential is fully compensated by the self-consistent electric potential. This compensation results in a spatially uniform
electron density. Therefore, there is no perturbation that would create excitations at the Fermi surface.

\begin{figure}[!ht]
\begin{center}
\includegraphics[width=0.4\textwidth,clip]{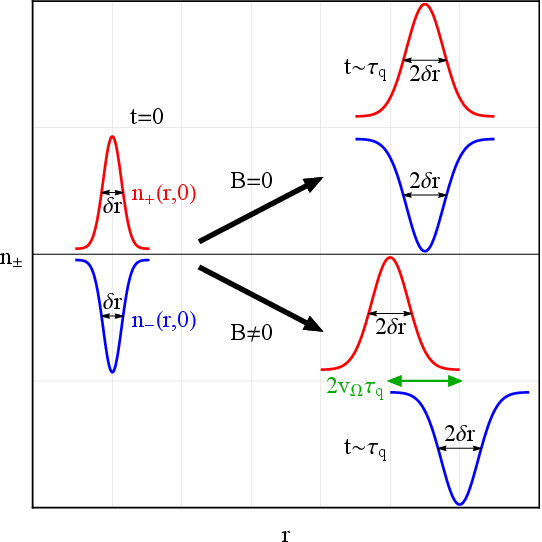}
\end{center}
\vspace{-0.5cm}
\caption{
Schematic illustration of the magnetic field effect on the electron quasiparticle density perturbations caused by the valley-dependent deformation potential. The initial spread of the density deviations $\delta r\sim 2\pi/q$ at $t=0$ increases to $\sim 2\delta r$ at $t\sim\tau_q=(Dq^2)^{-1}$. The magnetic field leads to the drift of the quasiparticles of opposite chiralities in opposite directions quantified by the anomalous velocity $v_{\Omega}$. This drift results in the spatially nonuniform electric charge density which triggers screening effects and reduces the sound attenuation.
}
\label{fig:CME-scheme}
\end{figure}

Strong screening requires the Thomas-Fermi length to be much shorter than the sound wavelength $2\pi/q$. This condition is easily satisfied in conventional and Weyl semimetals. The presence of several valleys in a semimetal allows for a new mechanism of the sound absorption if the deformation potential is valley dependent, even if it remains independent of momentum in each of the valleys and the screening is strong~\cite{Weinreich-White:1959,Gurevich-Efros:1963,Gantsevich-Gurevich:1967}. The reason is that the sound wave may create electron density perturbations of opposite signs in two valleys without violating the electric charge neutrality condition. In the case of Weyl semimetals, this would correspond to the chiral charge density. To illustrate this, we consider perturbations of densities of opposite signs, $n_{+}(r, 0)=-n_{-}(r, 0)=n(r,0)$, with the length scales $\delta r\sim 2\pi/q$ created at time $t=0$ in two valleys. See Fig.~\ref{fig:CME-scheme} for a schematic illustration. The initial spread of the densities would roughly double after time $t\sim\tau_q= (Dq^2)^{-1}$ elapses. (For brevity, we assume here that $q^2D \tau_5\gg1$.) In the absence of magnetic field, the densities in respective valleys would remain of opposite signs, $n_{+}(r,t)=-n_{-}(r,t)= n(r, t)$, at all times. Therefore, the neutrality condition remains satisfied, making the valley-dependent deformation potential effective for sound absorption.

Until now, there is no difference between conventional multivalley and Weyl semimetals. A qualitative difference appears with the application of a magnetic field. The classically weak magnetic field ($\omega_c\tau\ll 1$) does not affect the kinetic coefficients or relaxation times and, consequently, does not change the sound attenuation in a conventional semimetal~\cite{Abrikosov:book-1988}. In view of their relativisticlike spectrum and nontrivial topological properties, the effects of a magnetic field in Weyl semimetals are profoundly different. The combination of the magnetic field and the Berry curvature, which is opposite in the two valleys with different chiralities, results in the drift of the corresponding quasiparticle densities in opposite directions, $n_{\pm}(r,t)= \pm n(r\mp v_\Omega t, t)$. This leads to the CME current~\cite{Vilenkin:1980,Fukushima:2008} with a nonzero divergence and, as the result, to the spatially-nonuniform electric charge density $n_{+}(r,t)+n_{-}(r,t)\neq 0$. Therefore, the magnetic field provides a link between the seemingly independent electric and chiral charges. Of course, the nonuniform electric charge density is screened. As the result, the valley-imbalance perturbations are also suppressed. This is the reason for the sound attenuation decrease in the presence of a magnetic field. The drift becomes important if its effect exceeds the effect of spreading, i.e., at $v_\Omega\tau_q\gtrsim \delta r$, see Fig.~\ref{fig:CME-scheme}. In the opposite case of large spreading, $v_\Omega\tau_q\lesssim \delta r$, the overlap of the densities prevails, and the deviations from the charge neutrality are weak. The effect of the magnetic field on the sound attenuation is small in this regime. The drift-diffusion competition is clearly seen in the denominator of Eq.~(\ref{Q-B-Gamma-B_limit}).

Finally, let us provide intuitive symmetry arguments explaining the appearance of the sound dichroism. The dependence of the sound attenuation coefficient on a (pseudo)magnetic field can be described via terms $\propto\left(\mathbf{v}_{\Omega,\alpha}\cdot\mathbf{q}\right)$. Depending on the relevant symmetries, both odd and even or only even powers of such terms are allowed. Let us start with a pseudomagnetic sound dichroism. The pseudomagnetic field $\mathbf{B}_5$ breaks both symmetry between the Weyl nodes and crystal symmetries. Since $\mathbf{v}_{\Omega,5}\propto \mathbf{B}_5$ is a polar vector, $\left(\mathbf{v}_{\Omega,5}\cdot \mathbf{q}\right)$ is a scalar. Therefore the attenuation coefficient $\Gamma(\mathbf{B}_5,\omega, \hat{\mathbf{q}})$, which is a scalar, can contain odd powers of $\left(\mathbf{v}_{\Omega,5}\cdot \mathbf{q}\right)$. The magnetic sound dichroism relies on the different velocities and/or intra-node relaxation rates for Weyl nodes of opposite chiralities resulting in different diffusion constants. Notice that the corresponding difference $\delta D$ is a pseudoscalar. Therefore terms odd in $\delta D \left(\mathbf{v}_{\Omega}\cdot \mathbf{q}\right)$ are also allowed in the attenuation coefficient $\Gamma(\mathbf{B},\omega, \hat{\mathbf{q}})$. Hence the sound dichroism originates from certain symmetry properties of the semimetal activated by an external (pseudo)magnetic field.

\section{Summary}
\label{sec:Summary}

We investigated the anomalous sound attenuation in Weyl semimetals in external magnetic and pseudomagnetic fields. It was found that the nontrivial topology of Weyl semimetals activated by these fields has an unusual manifestation in the sound absorption. In addition to presenting these results, our work extends, corrects, and provides alternative derivation for some of the conclusions reached in the literature.

We develop an effective way of evaluating the sound attenuation coefficient, by directly relating it to the solution of a linearized kinetic equation, see Eqs.~(\ref{Q-Gamma-fin}), (\ref{Q-Gamma-fin-equal}), and (\ref{CKT-sol-n-chi}). Under rather natural assumptions regarding the symmetry of Weyl nodes, a compact form of the solutions to kinetic equations allows us to elucidate the dependence on the material parameters as well as the field strength and orientation; see Eqs.~(\ref{CKT-sol-B-n-sol-expl}), (\ref{CKT-sol-B5-n-sol-fin}), (\ref{CKT-sol-B-b0-n-exp-1-vF-2}), and (\ref{CKT-sol-B-b0-n-exp-5-vF}).

The expression for the sound attenuation coefficient for vanishing magnetic and pseudomagnetic fields agrees with that in multivalley semiconductors~\cite{Weinreich-White:1959,Gurevich-Efros:1963,Gantsevich-Gurevich:1967}. However, unlike the earlier studies in Ref.~\cite{Spivak:2016}, we found that sound attenuation coefficient is generically suppressed by the magnetic field; see Eqs.~(\ref{Q-B-Gamma-B}) and (\ref{Q-B-Gamma-B_limit}). Indeed, the chiral anomaly activated by this field necessarily leads to the deviations in electric charge density and, consequently, to the appearance of induced electric field. This electric field alone, however, cannot explain the suppression of the sound absorption. As we explicitly showed, the other important ingredient is a nonvanishing intra-node relaxation time $\tau$. The combination of finite $\tau$ and the electrostatic screening strongly affects the sound absorption and, contrary to Ref.~\cite{Spivak:2016}, leads to a negative sign of the anomalous part of the sound attenuation coefficient. These findings agree with the recent results presented in Ref.~\cite{Pesin-Ilan:2020}. Our estimates suggest that the suppression could be noticeable and reaches several percent of the attenuation coefficient at zero field for sufficiently strong magnetic fields; see Fig.~\ref{fig:Q-B-num-Gamma}. This finding clearly distinguishes Weyl semimetals from conventional multivalley conductors, where the effects of classically weak magnetic fields on the sound attenuation are negligible.

The nontrivial dependence on the magnetic field direction with respect to the sound wave vector appears when the Weyl nodes of opposite chiralities are nonsymmetric. In particular, different Fermi velocities and/or intra-node scattering amplitudes for the Weyl nodes of opposite chiralities make the dependence on the magnetic field nonmonotonic and lead to the magnetic sound dichroism, see Eq.~(\ref{Q-B-b0-Gamma-1-2}). This effect, however, is estimated to be weak. For optimal magnetic fields and small difference between the nodes, it reaches a few percent of the attenuation coefficient at zero field, see Fig.~\ref{fig:Q-B5-num-Gamma-close}. The magnetic sound dichroism for nonsymmetric Weyl nodes is one of our main results.

The sound absorption in an external strain-induced pseudomagnetic field, which is unique for a relativisticlike energy spectrum, is also nontrivial. Similar to the case of a magnetic field, the sound attenuation coefficient decreases in large pseudomagnetic fields. The dependence is, however, nonmonotonic and demonstrates the pseudomagnetic sound dichroism even when the Weyl nodes in the undeformed material are symmetric; see Eqs.~(\ref{Q-B5-Gamma-All}) and (\ref{Q-B5-Gamma-All-weak-B5-odd}). Like magnetic sound dichroism, its pseudomagnetic counterpart is estimated to be weak. The decrease in the sound attenuation coefficient and the sound dichroism in an external pseudomagnetic field is another major result of our study.

The proposed effects provide a way to probe the anomalous properties of Weyl semimetals and the effects of magnetic and strain-induced pseudomagnetic fields via sound absorption experiments. It would be especially interesting to see both the magnetic and pseudomagnetic sound dichroism by flipping the direction of the sound propagation with respect to the field, as well as to observe the reduction of the attenuation coefficient for the sound propagating along the field. As possible material candidates, transition metal monopnictides TaAs, TaP, NbAs, and NbP~\cite{Hasan-Huang:rev-2017}, the magnetic compound EuCd$_2$As$_2$~\cite{Soh-Boothroyd:2019,Ma-Shi:2019}, and SrSi$_2$~\cite{Huang-Hasan:2016,Singh-Bansil-SrSi2:2018} could be used. The magnetic material EuCd$_2$As$_2$ is a promising candidate for the investigation of the pseudomagnetic sound dichroism since it breaks the time-reversal symmetry and contains only a single pair of Weyl nodes. As for the magnetic sound dichroism, transition metal monopnictides could be useful since they contain two types of Weyl nodes with different Fermi velocities. The Weyl nodes of opposite chiralities are separated in energy in SrSi$_2$ also leading to nonsymmetric nodal parameters.

\begin{acknowledgments}
The authors acknowledge useful communications with E.~V.~Gorbar, D.~Pesin, H.~Rostami, and I.~A.~Shovkovy. This work is supported by NSF Garnt No.~DMR-2002275 (L.I.G.). P.O.S. acknowledges the support through the Yale Prize Postdoctoral Fellowship in Condensed Matter Theory.
\end{acknowledgments}

\appendix

\section{Energy dissipation rate}
\label{sec:app-0}

In this appendix, we calculate the energy dissipation rate required for the sound attenuation coefficient defined in Eq.~(\ref{Q-Gamma-def}). The energy dissipation rate is defined as
\begin{equation}
\label{app-0-Q-def}
Q = \frac{1}{T} \int_{0}^{T}dt \langle \frac{d}{dt} \hat{H}\rangle.
\end{equation}
Here, $T$ is the period of sound waves and $\hat{H}=\hat{H}_{ee}+\hat{H}_{ep}$ is the full Hamiltonian that includes the electron-electron $\hat{H}_{ee}$ and the electron-phonon $\hat{H}_{ep}$ parts. We use the standard definition of the ensemble averaging
\begin{equation}
\label{app-0-dt-H-def}
\langle \frac{d}{dt} \hat{H}\rangle = \mbox{Tr}\left\{\hat{\rho} \frac{\partial}{\partial t} \hat{H}\right\}.
\end{equation}
In the case of the sound attenuation, only $\hat{H}_{ep}$ explicitly depends on time. Therefore,
\begin{eqnarray}
\label{app-0-dt-H-aver}
&&\langle \partial_t\hat{H}\rangle = \langle \partial_t\hat{H}_{ep}\rangle
= \sum_{\alpha}^{N_{W}} \int d^3r \int \frac{d^3p_{\alpha}}{(2\pi \hbar)^3} \lambda_{ij}^{(\alpha)}(\mathbf{p}_{\alpha}) \nonumber\\ 
&&\times \left[\partial_tu_{ij}(t,\mathbf{r})\right]  f_{\alpha}(\mathbf{r},\mathbf{p}_{\alpha}),
\end{eqnarray}
where the sum runs over all $N_{W}$ Weyl nodes, the valley-dependent deformation potential is $\lambda_{ij}^{(\alpha)}(\mathbf{p}_{\alpha}) u_{ij}(t,\mathbf{r})$, $u_{ij}(t,\mathbf{r})=\left(\partial_i u_j +\partial_j u_i\right)/2$ is the strain tensor, and $\mathbf{u}(t,\mathbf{r})$ is the displacement vector.

The distribution function $f_{\alpha}(\mathbf{r},\mathbf{p}_{\alpha})=f_{\alpha}^{(0)}(\mathbf{p}_{\alpha})+\delta f_{\alpha}(\mathbf{r},\mathbf{p}_{\alpha})$ contains the time- and coordinate-independent equilibrium part $f_{\alpha}^{(0)}(\mathbf{p}_{\alpha})$ and the oscillating nonequilibrium correction determined by the deformation potential
\begin{equation}
\label{app-1-Model-df}
\delta f_{\alpha}(\mathbf{r},\mathbf{p}_{\alpha}) \approx \delta\left(\epsilon_{\alpha} +\chi_{\alpha} b_0 -\mu\right)n_{\alpha}(\mathbf{r},\mathbf{p}_{\alpha}).
\end{equation}
Here, $\mu$ is the equilibrium Fermi energy measured from the Weyl nodes and we assume that temperature is small compared with $\mu$. In addition, $\epsilon_{\alpha}=\epsilon_{\alpha}(p_{\alpha})$ is the energy dispersion relation and $b_0$ quantifies the separation between the Weyl nodes of opposite chiralities in energy.

Due to the time-averaging in Eq.~(\ref{app-0-Q-def}), only the nonequilibrium part of the distribution function $\delta f_{\alpha}(\mathbf{r},\mathbf{p}_{\alpha}) \propto \lambda_{ij}^{(\alpha)}(\mathbf{p}_{\alpha}) u_{ij}(t,\mathbf{r})$ contributes to the dissipation rate. The final expression for the energy dissipation rate is
\begin{eqnarray}
\label{app-0-Q-fin-fin}
\frac{Q}{V} &=& \sum_{\alpha}^{N_{W}} \frac{1}{2} \mbox{Re}\left\{ \int \frac{d^3p_{\alpha}}{(2\pi \hbar)^3} i\omega\left[\lambda_{ij}^{(\alpha)} u_{ij}\right]^{*} \delta f_{\alpha}(\mathbf{p}_{\alpha}) \right\} \nonumber\\
&=& \sum_{\alpha}^{N_{W}} \frac{\nu_{\alpha}(\mu)}{2} \mbox{Re}\left\{i\omega\left[\lambda_{ij}^{(\alpha)} u_{ij}\right]^{*}\overline{n_{\alpha}}\right\}.
\end{eqnarray}
Here, we used the plane-wave dependence for the displacement vector $\mathbf{u}(t,\mathbf{r})=\mathbf{u}_0e^{-i\omega t +i\mathbf{q}\cdot\mathbf{r}}$, where $\omega$ and $\mathbf{q}$ are the sound frequency and wave vector, respectively, $\mathbf{u}_0$ is the displacement magnitude, and $V$ is the spatial integration volume. In addition, we defined the averaged over the Fermi surface solution of the kinetic equations (at given $\mathbf{q}$) as $\overline{n_{\alpha}}$,
\begin{eqnarray}
\label{app-1-intra-bar-def}
\overline{n_{\alpha}} &=& \frac{1}{\nu_{\alpha}(\mu)}\int \frac{d^3p_{\alpha}}{(2\pi \hbar)^3} \Theta_{\alpha}(\mathbf{p}_{\alpha}) \delta\Big\{\epsilon_{\alpha}\left[1 +\frac{e}{c} \left(\bm{\Omega}_{\alpha}\cdot\mathbf{B}_{\alpha}\right)\right] \nonumber\\
&+& \chi_{\alpha} b_0-\mu\Big\} n_{\alpha}(\mathbf{p}_{\alpha}).
\end{eqnarray}
Here
\begin{eqnarray}
\label{app-1-cc-DOS}
\nu_{\alpha}(\mu) &=& \int \frac{d^3p_{\alpha}}{(2\pi \hbar)^3}\Theta_{\alpha}(\mathbf{p}_{\alpha}) \delta\Big\{\epsilon_{\alpha}\left[1 +\frac{e}{c} \left(\bm{\Omega}_{\alpha}\cdot\mathbf{B}_{\alpha}\right)\right] \nonumber\\
&+& \chi_{\alpha} b_0-\mu\Big\}
\end{eqnarray}
is the density of states (DOS) per Weyl node. For the sake of generality, we restored the renormalization of the phase-space volume $\Theta_{\alpha}(\mathbf{p}_{\alpha})=\left[1-e \left(\mathbf{B}_{\alpha}\cdot \mathbf{\Omega}_{\alpha}\right)/c\right]$ and the magnetic moment in the energy dispersion $\epsilon_{\alpha}\to\epsilon_{\alpha}\left[1+ e \left(\bm{\Omega}_{\alpha}\cdot\mathbf{B}_{\alpha}\right)/c\right]$ in Eqs.~(\ref{app-1-intra-bar-def}) and (\ref{app-1-cc-DOS}). Here, $\mathbf{B}_{\alpha}$ is an external (pseudo)magnetic field, and $\mathbf{\Omega}_{\alpha}=\mathbf{\Omega}_{\alpha}({\mathbf p}_{\alpha})$ is the Berry curvature defined in Eq.~(\ref{CKT-equation-Berry-monopole}). In the case of weak (nonquantizing) magnetic fields, the main contribution in the magnetic field dependence of the sound attenuation coefficient is given by the chiral anomaly; see the discussion at the end of Sec.~\ref{sec:Q-B}. Therefore we can neglect the magnetic moment in the energy dispersion and set $\Theta_{\alpha}(\mathbf{p}_{\alpha})\approx1$~\cite{Son-Spivak:2013,Spivak:2016}.

The dissipation rate in Eq.~(\ref{app-0-Q-fin-fin}) agrees with the result in Ref.~\cite{Gurevich-Pavlov:1971} if the external electric field in that paper is ignored. It is in agreement also with the general definition of the absorbed power in Ref.~\cite{Galperin-Kozub:rev-1979}.

\section{Collision integrals}
\label{sec:app-1}

In this appendix, we consider the collision integral on the right-hand side of the kinetic equation (\ref{CKT-equation-kinetic-equation}). It is convenient to split the integral into the intra- and inter-node parts that are discussed in Appendixes~\ref{sec:app-1-intra} and \ref{sec:app-1-inter}, respectively.

\subsection{Intra-node scattering}
\label{sec:app-1-intra}

By using the Fermi golden rule (see, e.g., Ref.~\cite{Abrikosov:book-1988}), the collision integral for the intra-node scattering processes is defined as
\begin{eqnarray}
\label{app-1-intra-1}
&&I_{\rm intra}\left[f_{\alpha}(\mathbf{p}_{\alpha})\right] = - \int \frac{d^3p_{\alpha}^{\prime}}{(2\pi \hbar)^3} \Theta_{\alpha}(\mathbf{p}_{\alpha}^{\prime}) \frac{2\pi}{\hbar} \left|A_{\alpha, \alpha}\right|^2 \nonumber\\
&&\times \delta\left[\tilde{\epsilon}_{\alpha}(\mathbf{p}_{\alpha})-\tilde{\epsilon}_{\alpha}(\mathbf{p}_{\alpha}^{\prime})\right] \left[f_{\alpha}(\mathbf{p}_{\alpha})-f_{\alpha}(\mathbf{p}_{\alpha}^{\prime})\right],
\end{eqnarray}
where $f_{\alpha}(\mathbf{p}_{\alpha})$ is the distribution function for the electron quasiparticles at the node $\alpha$ (we suppressed the explicit dependence on spatial coordinates) and $\left|A_{\alpha, \beta}\right|$ is the scattering amplitude between the Weyl nodes $\alpha$ and $\beta$. In the approximation where the magnetic moment and phase-space volume renormalization are neglected, the equilibrium distribution function depends only on the absolute value of momentum, i.e., $f_{\alpha}^{(0)}(\mathbf{p}_{\alpha})\approx f_{\alpha}^{(0)}(p_{\alpha})$, and is given by the standard Fermi-Dirac distribution. Then, we can use Eq.~(\ref{app-1-Model-df}) for the deviations from the equilibrium state $\delta f_{\alpha}(\mathbf{p}_{\alpha})$. This allows us to rewrite the intra-node collision integral as
\begin{widetext}
\begin{eqnarray}
\label{app-1-intra-2}
I_{\rm intra}\left[f_{\alpha}(\mathbf{p}_{\alpha})\right] &\approx& - \int \frac{d^3p_{\alpha}^{\prime}}{(2\pi \hbar)^3} \frac{2\pi}{\hbar} \left|A_{\alpha, \alpha}\right|^2 \delta\left[\epsilon_{\alpha}(p_{\alpha})-\epsilon_{\alpha}(p_{\alpha}^{\prime})\right] \left[\delta f_{\alpha}(\mathbf{p}_{\alpha})-\delta f_{\alpha}(\mathbf{p}_{\alpha}^{\prime}) \right] \nonumber\\
&=& -\frac{n(\mathbf{p}_{\alpha}) -\overline{n_{\alpha}}}{\tau_{\alpha}} \delta\left[\epsilon_{\alpha}(p_{\alpha})+\chi_{\alpha} b_0 -\mu\right].
\end{eqnarray}
\end{widetext}
Here, we introduced the intra-node relaxation time
\begin{eqnarray}
\label{app-1-intra-tau-def}
\frac{1}{\tau_{\alpha}} \equiv \frac{1}{\tau_{\alpha,\alpha}} &=& \int \frac{d^3p_{\alpha}}{(2\pi \hbar)^3} \frac{2\pi}{\hbar} \left|A_{\alpha, \alpha}\right|^2 \delta\left[\epsilon_{\alpha}(p_{\alpha})+\chi_{\alpha} b_0-\mu\right] \nonumber\\ 
&=&  \frac{2\pi}{\hbar} \left|A_{\alpha, \alpha}\right|^2 \nu_{\alpha}(\mu).
\end{eqnarray}

\subsection{Inter-node scattering}
\label{sec:app-1-inter}

Furthermore, we proceed to the inter-node collision integral $I_{\rm inter}\left[f_{\alpha}(\mathbf{p}_{\alpha})\right]$. Because the deformation potential depends on the chirality of the Weyl nodes (see the discussion in Sec.~\ref{sec:Model-0}), there are two types of contributions in the collision integral, i.e.,
\begin{equation}
\label{app-1-Icoll-inter-0}
I_{\rm inter}\left[f_{\alpha}(\mathbf{p}_{\alpha})\right]=\sum_{\beta\neq\alpha}^{N_{W}} \frac{|\chi_{\alpha}-\chi_{\beta}|}{2} I_{\alpha, \beta}^{(1)}+\sum_{\beta\neq\alpha}^{N_{W}} \frac{|\chi_{\alpha}+\chi_{\beta}|}{2} I_{\alpha, \beta}^{(2)},
\end{equation}
where $\sum_{\beta\neq\alpha}^{N_{W}}$ runs over all nodes excluding $\beta=\alpha$. Here, $I_{\alpha, \beta}^{(1)}$ corresponds to the scattering between the Weyl nodes of the opposite chiralities and $I_{\alpha, \beta}^{(2)}$ describes the transfer between the nodes of the same chirality $\chi_{\alpha}$. Let us begin with the former part,
\begin{widetext}
\begin{eqnarray}
\label{app-1-Icoll-inter-1}
I_{\alpha, \beta}^{(1)} &=& - \int \frac{d^3p_{\beta}}{(2\pi \hbar)^3}\Theta_{\beta}(\mathbf{p}_{\beta}) \frac{2\pi}{\hbar} \left|A_{\alpha, \beta}\right|^2 \delta\left[\tilde{\epsilon}_{\alpha}(p_{\alpha})-\tilde{\epsilon}_{\beta}(p_{\beta})\right] \left[f_{\alpha}(\mathbf{p}_{\alpha})-f_{\beta}(\mathbf{p}_{\beta})\right] \nonumber\\
&\approx& - \int \frac{d^3p_{\beta}}{(2\pi \hbar)^3} \frac{2\pi}{\hbar} \left|A_{\alpha, \beta}\right|^2 \delta\left[\epsilon_{\alpha}(p_{\alpha})-\epsilon_{\beta}(p_{\beta}) +2\chi_{\alpha}b_0 +\lambda_{ij}^{(\alpha)}u_{ij}-\lambda_{ij}^{(\beta)}u_{ij}\right] \nonumber\\
&\times&\left[\delta f_{\alpha}(\mathbf{p}_{\alpha})-\delta f_{\beta}(\mathbf{p}_{\beta}) + f_{\alpha}^{(0)}(p_{\alpha})-f_{\beta}^{(0)}(p_{\beta})\right]
\approx - \int \frac{d^3p_{\beta}}{(2\pi \hbar)^3} \frac{2\pi}{\hbar} \left|A_{\alpha, \beta}\right|^2 \delta\left[\epsilon_{\alpha}(p_{\alpha})-\epsilon_{\beta}(p_{\beta}) +2\chi_{\alpha}b_0\right] \nonumber\\
&\times&\left[\delta f_{\alpha}(\mathbf{p}_{\alpha})-\delta f_{\beta}(\mathbf{p}_{\beta}) - \left(\lambda_{ij}^{(\alpha)} -\lambda_{ij}^{(\beta)}\right) u_{ij}(\partial_{\epsilon_{\beta}}f_{\beta}^{(0)})\right],
\end{eqnarray}
\end{widetext}
where we used
\begin{widetext}
\begin{eqnarray}
\label{app-1-Icoll-delta-1}
&&\delta\left(\epsilon_{\alpha}(p_{\alpha})-\epsilon_{\beta}(p_{\beta}) +2\chi_{\alpha}b_0 +\lambda_{ij}^{(\alpha)}u_{ij}-\lambda_{ij}^{(\beta)}u_{ij}\right) \left[f_{\alpha}^{(0)}(p_{\alpha})-f_{\beta}^{(0)}(p_{\beta})\right] \nonumber\\
&&\approx \delta\left[\epsilon_{\alpha}(p_{\alpha})-\epsilon_{\beta}(p_{\beta}) +2\chi_{\alpha}b_0\right]\left[f_{\alpha}^{(0)}(p_{\alpha})-f_{\beta}^{(0)}(p_{\alpha}) -\left(\lambda_{ij}^{(\alpha)} -\lambda_{ij}^{(\beta)}\right) u_{ij}(\partial_{\epsilon_{\beta}}f_{\beta}^{(0)})\right]
\end{eqnarray}
\end{widetext}
and neglected the second-order terms $\propto (\lambda_{ij}^{(\alpha)}u_{ij})^2$ in the last line in Eq.~(\ref{app-1-Icoll-inter-1}).

By introducing the inter-node relaxation time:
\begin{eqnarray}
\label{app-1-intra-tau-inter-def}
\frac{1}{\tau_{\alpha, \beta}} &=& \int \frac{d^3p_{\beta}}{(2\pi \hbar)^3} \frac{2\pi}{\hbar} \left|A_{\alpha, \beta}\right|^2 \delta\left[\epsilon_{\beta}(p_{\beta}) +\chi_{\beta} b_0-\mu\right] \nonumber\\
&=& \frac{2\pi}{\hbar} \left|A_{\alpha, \beta}\right|^2 \nu_{\beta}(\mu),
\end{eqnarray}
we rewrite Eq.~(\ref{app-1-Icoll-inter-1}) as
\begin{eqnarray}
\label{app-1-Icoll-inter-1-fin}
I_{\alpha, \beta}^{(1)} &=& -\frac{n_{\alpha}(\mathbf{p}_{\alpha}) -\overline{n_{\beta}}}{\tau_{\alpha, \beta}} \delta\left(\epsilon_{\alpha}+\chi_{\alpha} b_0-\mu\right) \nonumber\\
&-& \frac{\left(\lambda_{ij}^{(\alpha)} -\lambda_{ij}^{(\beta)}\right)u_{ij}}{\tau_{\alpha, \beta}} \delta\left(\epsilon_{\alpha}+\chi_{\alpha} b_0-\mu\right).
\end{eqnarray}

The result for the inter-node scattering integral between the nodes of the same chirality can be obtained in the same way. It reads as
\begin{equation}
\label{app-1-Icoll-inter-2-fin}
I_{\alpha, \beta}^{(2)} = -\frac{n_{\alpha}(\mathbf{p}_{\alpha}) -\overline{n_{\beta}}}{\tau_{\alpha, \beta}} \delta\left(\epsilon_{\alpha}+\chi_{\alpha} b_0-\mu\right).
\end{equation}

Then, the inter-node collision integral in the $\tau$ approximation is
\begin{eqnarray}
\label{app-1-Icoll-inter-fin}
&&I_{\rm inter}\left[f_{\alpha}(\mathbf{p}_{\alpha})\right] = -\sum_{\beta\neq\alpha}^{N_{W}}\frac{n_{\alpha}(\mathbf{p}_{\alpha}) -\overline{n_{\beta}}}{\tau_{\alpha,\beta}} \delta\left(\epsilon_{\alpha}+\chi_{\alpha}b_0-\mu\right) \nonumber\\
&&-  u_{ij}\sum_{\beta}^{N_{W}} \frac{\lambda_{ij}^{(\alpha)} -\lambda_{ij}^{(\beta)}}{\tau_{\alpha,\beta}}  \delta\left(\epsilon_{\alpha}+\chi_{\alpha}b_0-\mu\right).
\end{eqnarray}
After averaging over the Fermi surface, one obtains
\begin{equation}
\label{app-1-Icoll-inter-fin-aver}
\overline{I_{\rm inter}\left[f_{\alpha}(\mathbf{p}_{\alpha})\right]} = -\sum_{\beta}^{N_{W}}\frac{\overline{n_{\alpha}} -\overline{n_{\beta}}}{\tau_{\alpha,\beta}} - u_{ij}\sum_{\beta}^{N_{W}}  \frac{\lambda_{ij}^{(\alpha)} -\lambda_{ij}^{(\beta)}}{\tau_{\alpha,\beta}}.
\end{equation}
This collision integral is used in Eq.~(\ref{CKT-equation-n-chi}) in the main text.

For the deformation potential that depends only on the chirality of Weyl nodes, i.e.,  $\lambda_{ij}^{(\alpha)}=\lambda_{ij}+\chi_{\alpha}\lambda_{ij}^{(5)}$, it is convenient to introduce the following inter-node scattering time for the Weyl nodes of opposite chiralities:
\begin{equation}
\label{app-1-Icoll-inter-tau5-def}
\frac{1}{\tau_{5,\alpha}} = \sum_{\beta}^{N_{W}} \frac{|\chi_{\alpha}-\chi_{\beta}|}{\tau_{\alpha,\beta}}.
\end{equation}

\bibliography{library_attenuation}

\end{document}